\documentclass[journal,11pt,onecolumn]{IEEEtran}

\IEEEoverridecommandlockouts                              

\usepackage{times}
\usepackage{amsthm}
\usepackage{amsmath}    
\usepackage{amssymb}    
\usepackage{mathrsfs}   
\usepackage{epsfig}     
\usepackage{graphicx}

\usepackage{comment}

\usepackage{subfigure}

\usepackage{amsfonts}
\usepackage{dsfont}

\usepackage{multirow}

\usepackage{cite}

\usepackage{enumerate}

\newcommand{\ep}{\epsilon}

\newcommand{\A}{{\mathcal A}}
\newcommand{\B}{{\mathcal B}}
\newcommand{\C}{{\mathcal C}}

\renewcommand{\H}{{\mathbf H}}

\newcommand{\E}{{\mathcal E}}
\newcommand{\V}{{\mathcal V}}
\renewcommand{\S}{{\mathcal S}}
\newcommand{\I}{{\mathcal I}}
\newcommand{\M}{{\mathcal M}}
\newcommand{\N}{{\mathcal N}}
\newcommand{\U}{{\mathbf U}}

\newcommand{\dE}{D_\Sigma}

\newcounter{constcount}
\newcommand{\eref}[1]{(\ref{#1})}

\newcounter{numcount}
\newcommand{\Nmimo}{\N_{\rm BC}}
\newcommand{\Hmimo}{\H_{\rm BC}}

\newcounter{thmcnt}
  \let\Oldsection\section
  
\renewcommand{\section}{\stepcounter{thmcnt}\Oldsection}

\allowdisplaybreaks

\newtheorem{theorem}{Theorem} 
\newtheorem{lemma}{Lemma} 
\newtheorem{definition}{Definition} 
\newtheorem{claim}{Claim} 

\newtheorem{remark}{Remark} 
\newtheorem{question}{Question}[section]

\newcounter{examplecounter}
\newcommand{\aln}[1]{\begin{align*}#1\end{align*}}

\newcommand{\al}[1]{\begin{align}#1\end{align}}

\def\Item$#1${\item $\displaystyle#1$
   \hfill\refstepcounter{equation}(\theequation)}

\newcommand{\bea}{\begin{eqnarray}}
\newcommand{\eea}{\end{eqnarray}}
\newcommand{\beas}{\begin{eqnarray*}}
\newcommand{\eeas}{\end{eqnarray*}}

\begin{document}

\title{On the Degrees-of-Freedom of Two-Unicast Wireless Networks with Delayed CSIT}

\author{Alireza Vahid
\thanks{Alireza Vahid is with the Electrical Engineering Department of the University of Colorado Denver {\tt\footnotesize alireza.vahid@ucdenver.edu}}
\thanks{The preliminary results of this work were presented at the 53rd Annual Allerton Conference on Communication, Control, and Computing (Allerton)~\cite{vahid2015informational}.}
}

 


\maketitle
\thispagestyle{empty}
\pagestyle{empty}

\begin{abstract}
We characterize the degrees-of-freedom (DoF) region of a class of two-unicast wireless networks under the assumption of delayed channel state information at the transmitters. We consider a layered topology with \emph{arbitrary} connectivity, and we introduce new outer-bounds on the DoF region through the graph-theoretic notion of bottleneck nodes. Such nodes act as informational bottlenecks \emph{only} under the assumption of delayed channel state information. Combining our outer-bounds with new achievability schemes, we characterize the DoF region of two-unicast wireless networks with informational bottlenecks. We show that unlike the instantaneous channel state information model, the sum DoF of two-unicast networks with delayed channel knowledge can take an \emph{infinite} set of values. We compare our results to the best previously known outer-bounds, and we show that the gap can be arbitrary large in favor of the current work.
\end{abstract}



\section{Introduction}
\label{Section:Introduction}

The classical result of Ford and Fulkerson~\cite{FF56} establishes the capacity of single-unicast wireline networks. Many extensions of this result are known today. In particular, single-flow networks are well-understood and known to obey max-flow min-cut type principles, both for the case of wireline networks~\cite{ACLY00} and for the case of wireless networks~\cite{ADT11}. However, obtaining capacity results for multi-flow networks seems to be a distant goal.

As a natural first step in studying multi-flow problems, networks with two source-destination pairs, or \emph{two-unicast} networks, have recently been the focus of significant attention~\cite{Ness2unicast,ShenviDey,xx,WangTwoUnicast,dof2unicastfull,zeng2015alignment}. But as it happens, characterizing the capacity of two-unicast wireline networks is as hard as the general $k$-unicast wireline problem~\cite{TwoUnicastIsHard}. In the wireless world, matters become even more challenging since signals transmitted at different nodes interfere with each other, causing the two information flows to mix.

In an attempt to obtain first-order capacity approximations and to capture the impact of interference in multi-flow wireless networks, a number of recent papers have focused on characterizing the degrees-of-freedom (DoF) of different network configurations. In essence, the DoF of a wireless network measures the pre-log factor in the capacity expression, and can be thought of as the gain over time-sharing. As a result of DoF studies, several new interference management techniques have recently been introduced, and shown to provide significant performance gains over simple time-sharing approaches~\cite{CadambeJafar,MotahariRealInterference,xx,dofkkk}. In particular, a careful combination of interference avoidance, interference neutralization, interference alignment~\cite{CadambeJafar,MotahariRealInterference} and aligned interference neutralization~\cite{xx} was used in~\cite{dof2unicastfull} to characterize the DoF of two-unicast layered wireless networks under the assumption of instantaneous channel state information (CSI) at all wireless nodes. However, as wireless networks grow in size, nodes turn mobile, and fast-fading channels become ubiquitous, providing instantaneous CSI is practically infeasible. In such scenarios, a more realistic model is the delayed channel state information at the transmitters (CSIT) in which by the time the CSI arrives at the transmitters, the channel has already changed to a new state.


In this work we study the impact of delayed CSIT in multi-hop multi-flow wireless networks by focusing our attention on two-unicast layered networks with arbitrary connectivity. In the case of instantaneous CSIT, it is known that the sum DoF of these networks can only take the values $1, 3/2,$ and $2$, and can be determined based on two graph-theoretic structures~\cite{dof2unicastfull}: the first one is the notion of paths with manageable interference, which captures when the two information flows can coexist and achieve a total of $2$ sum DoF, and the second one is the notion of an \emph{omniscient node}, which creates an informational bottleneck and limits the DoF to $1$. Whenever neither of these structures is found, the DoF is limited to $3/2$.

The case of delayed CSIT was previously considered in~\cite{WangTwoUnicastDCSIT}. Interestingly, it was shown that as long as no omniscient node is found, at least $4/3$ DoF is achievable. Hence, just as in the instantaneous CSIT case, the omniscient node is the key informational bottleneck whose absence determines when we can go beyond $1$ DoF (corresponding to a simple time-sharing scheme). However, it is also known that unlike the instantaneous CSIT case, networks with delayed CSIT may have $4/3$ DoF. Two questions arise: 1) how much richer is the set of possible DoF values in the delayed CSIT case? and 2) what are the new informational bottleneck structures that apply only to the case of delayed CSIT?
 
In this paper, we provide answers to both questions. First, we generalize the concept of an omniscient node and introduce the notion of a $\rho$-bottleneck node, $\rho \in \mathbb{N}$. Roughly speaking, all information flows to a destination, say $d_1$, have to pass through a $\rho$-bottleneck node. Moreover, the bottleneck node has a subset of parent nodes, $\M$, that all information flows from source $s_2$ have to pass through. The rank of the transfer matrix from $\M$ to the following layer is (almost surely) $\rho$. When a two-unicast network contains such a node for destination $d_i$, the DoF region is governed by 
\begin{align}
\label{Eq:NewRankBounds}
\rho D_i + D_{\bar{i}} \leq \rho, \qquad i=1,2,
\end{align}
where $D_i$ is the DoF for source-destination pair $i$, and $\bar{i} = 3 - i$. Second, we show that there exist two-unicast networks containing $\rho$-bottleneck nodes in which these outer-bounds are tight. We show that unlike several recent DoF characterizations where the sum DoF only attain a small and finite set of values~\cite{dof2unicastfull,XieTwoUnicastSecureDoF,WangX2Unicast}, the set of DoF values for two-unicast networks with delayed CSIT is in fact \emph{infinite}. More precisely, we show that there exist two-unicast layered networks with delayed CSIT and sum DoF taking any value in the set 
\begin{align} 
\S \triangleq \left\{ 2\left(1 - \frac{1}{k} \right) : k=1,2,... \right\}  \cup \left\{ 2 \right\}. 
\end{align}


In~\cite{vahid2015informational} we introduced the notion of $|\M|$-bottleneck nodes where $|\M|$ is the size of the subset of parent nodes introduced above for the bottleneck node, and we provided outer-bounds of the form $|\M| D_1 + D_2 \leq |\M|$. In this work we construct two-unicast networks in which the outer-bounds of~\cite{vahid2015informational} become loose and we demonstrate intuitively why instead of the size of $\M$, we should consider the rank of the transfer matrix from this set to the layer containing the bottleneck node, and we denote this rank by $\rho$. Since the rank of a matrix is less than or equal to the number of its columns, the outer-bounds in (\ref{Eq:NewRankBounds}) are tighter\footnote{If the rank equals the number of columns, then the two sets of bounds are identical.} than the ones given in~\cite{vahid2015informational}. In fact, in Section~\ref{Section:Compare} we construct a class of networks in which $|\M| \rightarrow \infty$ but  $\rho$ remains constant. As a result, the outer-bounds of~\cite{vahid2015informational} reduce to trivial bounds $D_i \leq 1$. However, the outer-bounds in (\ref{Eq:NewRankBounds}) remain unchanged and active.

A natural follow-up question is whether the new outer-bounds in (\ref{Eq:NewRankBounds}) suffice to characterize the DoF region of two-unicast layered networks with delayed CSIT. The answer to this question is negative. In particular, we present a two-unicast layered network in Section~\ref{Section:Discussion} for which the DoF under delayed CSIT is given by
\begin{equation}
\left\{ \begin{array}{ll}
\vspace{1mm} 0 \leq D_i \leq 1, & i = 1,2, \\
D_1 + D_2 \leq \frac{3}{2}. &
\end{array} \right.
\end{equation}
This region cannot be expressed using the outer-bounds in (\ref{Eq:NewRankBounds}). Moreover, unlike the other networks we consider in this paper, to achieve this DoF region our achievability strategy goes over infinitely many time slots. We provide a detailed discussion in Section~\ref{Section:Discussion}.

The paper is organized as follow. In Section~\ref{Section:Problem} we introduce the problem setting and our assumptions. We present our contributions in Section~\ref{Section:Main} followed by a number of motivating examples in Section~\ref{Section:Examples}. We formally define the notion of $\rho$-bottleneck nodes in Section~\ref{Section:Bottleneck} and prove the outer-bounds given in (\ref{Eq:NewRankBounds}). We then prove our main results in Section~\ref{Section:Proof}. We provide some further insights in Section~\ref{Section:Conclusion}, and conclude the paper in Section~\ref{Section:Conclusion}.


\section{Problem Setting}
\label{Section:Problem}

A multi-unicast (Gaussian) wireless network $\N = (G,L)$ consists of a directed graph $G=(\V,\E)$, where $\V$ is the node set and $\E \subset \V \times \V$ is the edge set, and a set of  source-destination pairs $L \subset \V \times \V$. In this work we focus on two-unicast Gaussian networks, \emph{i.e.} $L = \{ (s_1,d_1),(s_2,d_2)\}$, for distinct vertices $s_1, s_2, d_1, d_2 \in \V$. Moreover, we assume that the network is layered, meaning that the vertex set $\V$ can be partitioned into $r$ subsets $\V_1,\V_2,...,\V_r$ (called layers) in such a way that 
\begin{align}
\V_1 = \{s_1,s_2\}, \qquad \V_r = \{d_1,d_2\}, \qquad \E \subseteq \bigcup_{i=1}^{r-1} \V_i \times \V_{i+1}.  
\end{align}

For a vertex $v \in \V_j$, $j=2,3,\ldots,r$, we define the set of parent nodes of $v$ as 
\begin{align}
\I(v) \triangleq \{u\in V_{j-1}: (u,v) \in \E \}.
\end{align} 
 
A real-valued channel gain $h_{ji}[t]$ is associated with the edge from $v_i$ to $v_j$ at each time $t$. We consider a fast-fading scenario in which channel gains $\{h_{ji}[t]\}_{t=1}^\infty$ are assumed to be mutually independent random processes obeying an absolutely continuous distribution with finite variance. At time $t = 1,2,\ldots,n$, each node $v_i \in \V$ transmits a real-valued signal $X_{v_i}[t]$, which must satisfy an average power constraint 
\begin{align}
\frac1n \sum_{t=1}^n E\left[X_{v_i}^2[t]\right] \leq P, \qquad \forall v_i \in \V, 
\end{align}
for a communication block of length $n$. The signal received by node $v_j$ at time $t$ is given by
\begin{align} 
\label{receivedsignaldef}
Y_{v_j}[t] = \sum_{v_i \in \I(v_j)} h_{ji}[t] X_i[t] + Z_j[t], 
\end{align}
where $Z_j[t]$ is a zero-mean unit-variance Gaussian noise at node $v_j$, assumed to be i.i.d. across time and across nodes. We use $X_{v_i}^n$ to represent the vector $(X_{v_i}[1],...,X_{v_i}[n])$. For a subset of nodes $\A$, we define $X_{\A}[t] \overset{\triangle}= \left\{ X_{v_i}[t] : v_i \in \A \right\}$.


We consider a delayed CSIT model in which instantaneous knowledge of a channel gain realization is only available at the receiver end of that channel, and is learned with a unit delay at all other nodes. More precisely, we assume that at time $t$, a node $v_k \in \V$ has knowledge of 
\aln{
\{ h_{ki}^t : v_i \in \I(k)\} \cup \{ h_{k'i'}^{t-1} : (i',k') \in \E \}.
}

We use $\H^t = \left( h_{ji}[\ell] : (i,j) \in \E, 1 \leq \ell \leq t \right)$ to denote the random vector corresponding to the channel state information up to time $t$. We point out that other more restrictive delayed CSIT models where nodes learn channel gains with a longer delay, or with a delay that is proportional to how far a given channel is in the network~\cite{aggarwal2011achieving,WangTwoUnicastDCSIT,vahid2017interference} can be considered. However, it is straightforward to see that, through an interleaving operation, such models can be reduced to the model considered here.

We will use standard definitions for a coding scheme, an achievable rate pair $(R_1,R_2)$, and the capacity region $\mathcal{C}(P)$ of a network $\N$.
We say that the DoF pair $(D_1,D_2)$ is achievable if we can find achievable rate pairs $(R_1(P),R_2(P))$ such that 
\begin{align}
\label{Eq:DefinitionDoF}
D_i = \lim_{P\rightarrow \infty}\frac{R_i(P)}{\frac12 \log P}.
\end{align}
The DoF region $\mathcal{D}$ is defined as the closure of all achievable DoF pairs $(D_1,D_2)$. Moreover, the sum DoF, $\dE$, is defined as the supremum of $D_1+D_2$ over all achievable DoF pairs $(D_1,D_2)$.


\section{Main Results}
\label{Section:Main}

Recent results on the DoF characterization of multi-flow networks reveal a similar phenomenon: for (Lebesgue) almost all values of channel gains, the sum DoF is restricted to a small \emph{finite} set of values. In~\cite{dof2unicastfull} it is shown that $\dE \in \{1,3/2,2\}$ for two-unicast layered networks. When the secure DoF of two-unicast is considered instead,~\cite{XieTwoUnicastSecureDoF} shows that we must have $\dE \in \{0,2/3,1,3/2,2\}$. In~\cite{WangX2Unicast} two-source two-destination networks with arbitrary traffic demands were instead considered, and the set of sum DoF values was shown to be $\{1,4/3,3/2,2\}$. Finally, for the delayed CSIT setting considered in this paper, the authors in~\cite{WangTwoUnicastDCSIT} show that if $\dE \ne 1$, then $\dE \geq 4/3$, suggesting that perhaps in this case, $\dE$ is also restricted to a small number of discrete values.

In~\cite{vahid2015informational} we proved that this in not case and the sum DoF of two-unicast wireless networks with delayed CSIT can take infinitely many values. In this paper we improve upon the results of~\cite{vahid2015informational} by providing new tighter outer-bounds. In this section we provide a set of possible sum DoF values for two-unicast layered networks with delayed CSIT. To prove this result we will need new outer-bounds that are provided in Section~\ref{Section:Bottleneck}.

\begin{theorem} \label{mainthm}
There exist two-unicast layered networks with delayed CSIT and sum DoF, $\dE$, taking any value in the set\footnote{For $k = 1$ the sum DoF in (\ref{Seq}) is zero which corresponds to a degenerate two-unicast network.} 
\al{ \label{Seq}
\S \triangleq \left\{ 2\left(1 - \frac{1}{k} \right) : k=1,2,... \right\}  \cup \left\{ 2 \right\}. 
}
\end{theorem}
\vspace{2mm}

\begin{remark}
The statement of this theorem is identical to Theorem~1 of~\cite{vahid2015informational}. In Section~\ref{Section:Compare} we construct a class of networks in which the results of~\cite{vahid2015informational} imply trivial outer-bounds $D_i \leq 1$. On the other hand, we show that in these networks tighter outer-bounds given in Section~\ref{Section:Bottleneck} remain active. Moreover, as we will see in Theorem~\ref{bottlethm}, the gap between the results of the two papers can be arbitrary large. 
\end{remark}

Intuitively, the reason why the sum DoF of two-unicast wireless networks can take all values in $\S$ is the fact that the delayed CSIT setting creates new informational bottlenecks in the network. In this work, we identify a class of such structures, which we term $\rho$-bottleneck nodes ($\rho \in \mathbb{N}$). We defer the formal definition of an $\rho$-bottleneck node to Section~\ref{Section:Bottleneck}, but we describe its significance with an example.
\begin{figure}[t]
\centering
\includegraphics[height = 1.5in]{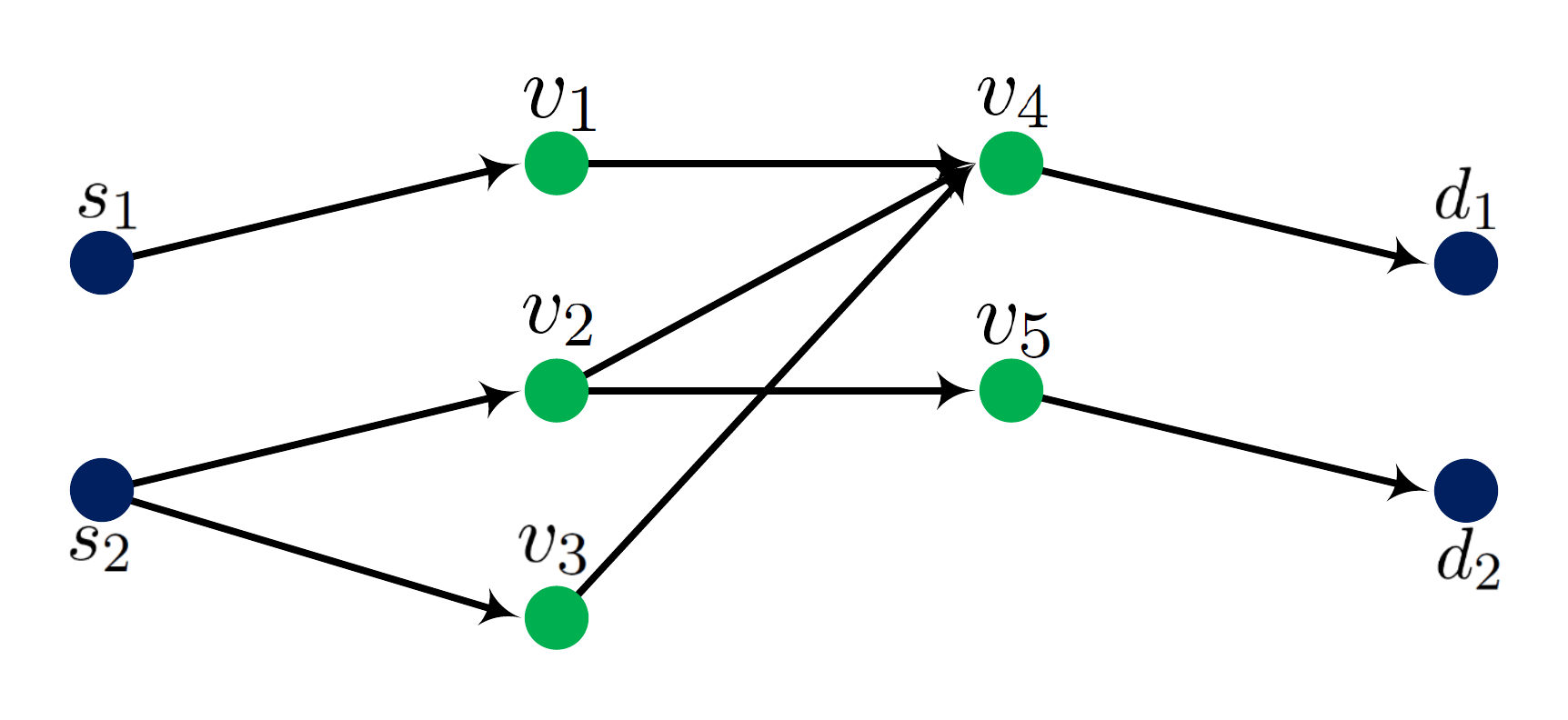}
\caption{Example of a network containing a bottleneck node ($v_4$).}\label{Fig:2D1D2}
\end{figure}
Consider the network in Fig.~\ref{Fig:2D1D2}. If instantaneous CSIT were available, $v_2$ and $v_3$ could amplify-and-forward their received signals with carefully chosen coefficients so that their signals cancel each other at receiver $v_4$. This would effectively create an interference-free network, and the cut-set bound of $2$ DoF would be achievable. However, when only delayed CSIT is available, such an approach is no longer possible. In fact, as we show in Section~\ref{Section:Bottleneck}, $v_4$ functions as a $2$-bottleneck node for destination $d_1$, causing the DoF to be constrained by
\aln{
2D_1 + D_2 \leq 2.
}
As it turns out, by utilizing delayed CSIT, the DoF pair $(1/2,1)$ can in fact be achieved.

In general, we show that whenever a network contains a $\rho$-bottleneck node for destination $d_i$ under the delayed CSIT assumption, we have
\al{ \label{bottlebound}
\rho D_{i} + D_{\bar{i}} \leq \rho,
}
where we let $\bar{i} = 3 - i$, and $i =1,2$. We point out that for $\rho=1$ a bottleneck node reduces to an omniscient node~\cite{WangTwoUnicast,ilanthesis,WangTwoUnicastDCSIT} which was known to be an informational bottleneck in two-unicast networks, even under instantaneous CSIT.

It is now a good time to highlight the difference between the current paper and~\cite{vahid2015informational} in more detail. For the outer-bounds in~\cite{vahid2015informational}, \emph{i.e.}
\begin{align}
|\M| D_{i} + D_{\bar{i}} \leq |\M|, \qquad i=1,2,
\end{align}
$|.|$ denotes the size of a set, and $\M$ is a set of parent nodes of the bottleneck node through which all information flows from source $s_{\bar{i}}$ have to pass. However, in the current results, $\rho$ is the rank of the transfer matrix from $\M$ to the following layer, and thus the outer-bounds in this paper are tighter when compared to~\cite{vahid2015informational}. In fact, in Section~\ref{Section:Compare}, we construct an example to highlight the gap between the two results.

In addition, we show that it is possible to build a two-unicast layered network where the outer-bound implied by~\eref{bottlebound} is tight. In order to do so, we introduce linear achievability schemes that make use of delayed CSIT in order to reduce the effective interference experienced by the bottleneck nodes as much as possible. Theorem~\ref{mainthm} then follows by noticing that if we have a network with a $\rho$-bottleneck node for $d_1$ and an $\rho$-bottleneck node for $d_2$, then we must have $\rho D_1 + D_2 \leq \rho$ and $D_1+ \rho D_2 \leq \rho$, which implies
\aln{
(\rho+1) (D_1 + D_2) \leq 2\rho \; \Rightarrow \;  D_1 + D_2 \leq 2 \left( 1 - 1/(\rho+1) \right).
}
Showing that two-unicast networks exist where the bound above is tight implies Theorem~\ref{mainthm}. Before proving our main results, we present two motivating examples to describe the role of an $\rho$-bottleneck node.


\section{Motivating Examples}
\label{Section:Examples}

In this section we illustrate the concept of a bottleneck node through two examples, and we illustrate the transmission strategies that take advantage of delayed CSIT. These examples are borrowed from~\cite{vahid2015informational}. We later investigate a third example in Section~\ref{Section:Compare} that reveals the shortcoming of the results in~\cite{vahid2015informational} and demonstrates intuitively why instead of the size of $\M$, we should consider the rank of the transfer matrix from this set to the layer containing the bottleneck node. We formally define the notion of $\rho$-bottleneck nodes in Section~\ref{Section:Bottleneck}.


\subsection{Example~1: A Two-Unicast Network with a Bottleneck Node}
\label{bottleexsec}

\begin{figure}[t]
\centering
\includegraphics[height = 1.5in]{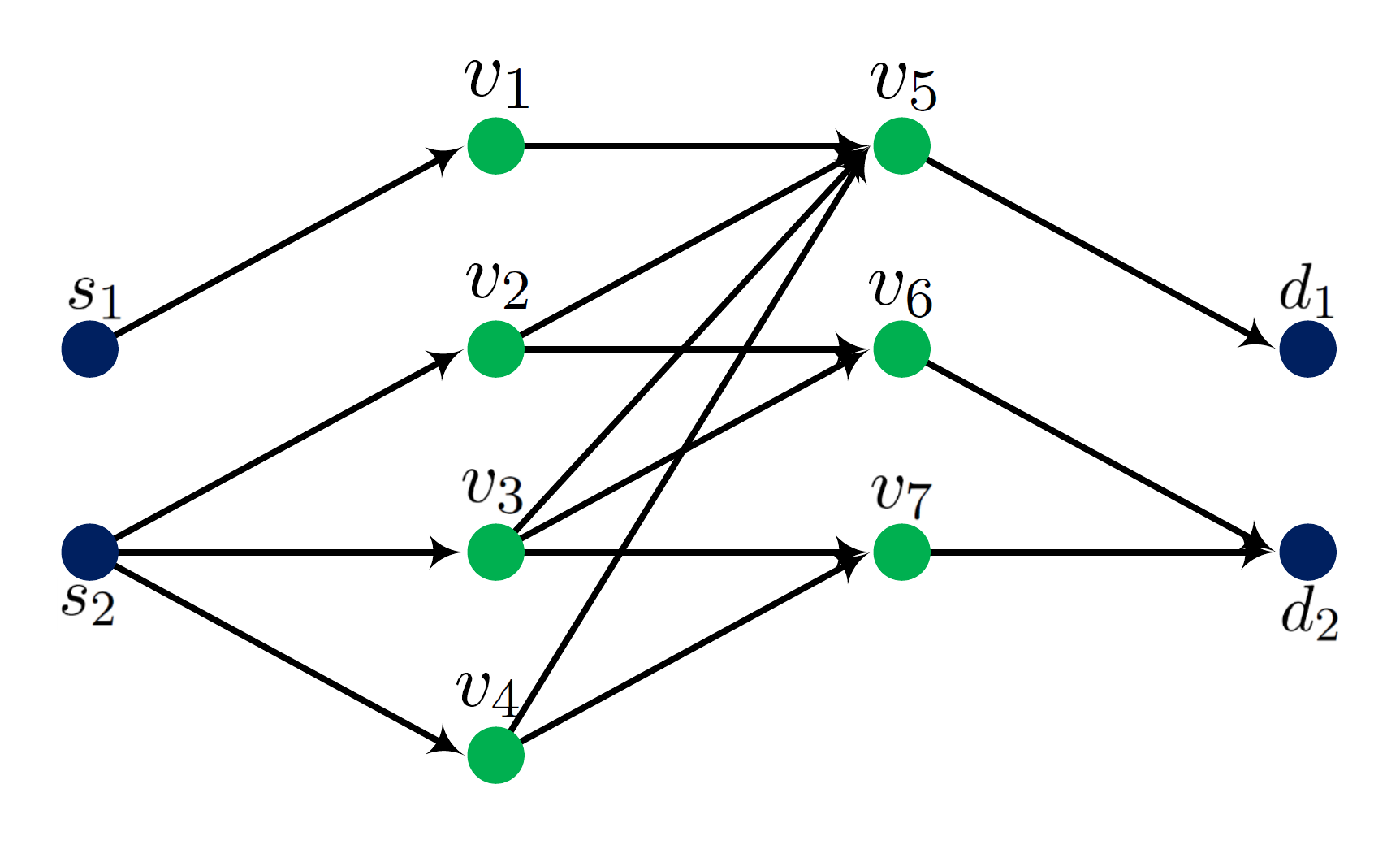}
\caption{Motivating example: we show that for this network, using the delayed CSIT, we can achieve $\left( D_1, D_2 \right) = \left( 2/3, 1 \right)$. $v_5$ acts as an informational bottleneck node in this network.}\label{Fig:3D1D2}
\end{figure}

Consider the network depicted in Fig.~\ref{Fig:3D1D2}. If instantaneous CSIT was available, $v_2, v_3$ and $v_4$ could scale their signals such that their interference at $v_5$ is canceled. However, when CSIT is only available with delay, such an approach does not work, and in order for information to flow from $s_2$ to $d_2$, some interference inevitably occurs at $v_5$. This suggests that $v_5$ plays the role of an informational bottleneck and the sum DoF should be strictly smaller than $2$.

We show that for this network we can achieve $\left( D_1, D_2 \right) = \left( 2/3, 1 \right)$. To do so, it suffices to show that during three time slots source $s_1$ can communicate two symbols to destination $d_1$, while source $s_2$ can communicate three symbols to destination $d_2$.
Since we can concatenate many three-slot communication blocks, we can describe our encoding as if the three time slots for the first hop occur first, followed by the three time slots for the second hop, and finally, the time slots for the third hop. By concatenating many blocks, the delay from waiting three time slots at each layer becomes negligible.  Next, we describe the transmission strategy for each hop separately. This way, there will be no issues regarding causality in the network. We will ignore noise terms to simplify the exposition in this section.

\vspace{2mm}

\noindent {\bf Transmission strategy for the first hop:} During the first two time slots each source sends out two symbols: source $s_1$ sends out symbols $a_1$ and $a_2$, while source $s_2$ sends out symbols $b_1$ and $b_2$. During the third time slot, source $s_1$ remains silent while source $s_2$ sends out one symbol denoted by $b_3$. We note that upon completion of these three time slots, relay $v_1$ has access to symbols $a_1$ and $a_2$, and relay $v_j$ has access to symbols $b_1$, $b_2$, and $b_3$, $j=2,3,4$.

\vspace{2mm}

\begin{figure}[ht]
\centering
\includegraphics[height = 2.5in]{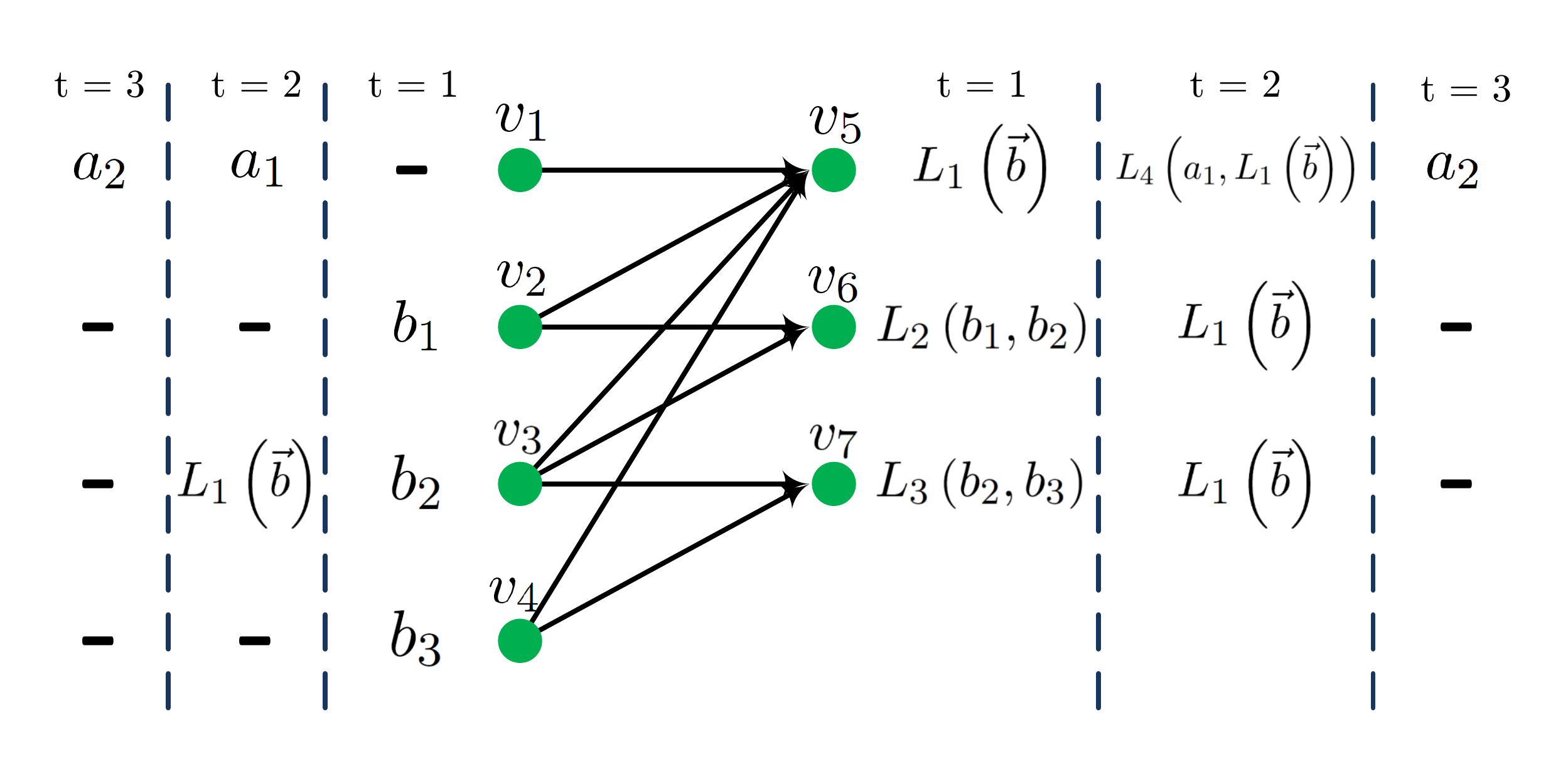}
\caption{Transmission strategy for the second hop of the network depicted in Fig.~\ref{Fig:3D1D2}.}\label{Fig:SecondHopEx1}
\end{figure}

\noindent {\bf Transmission strategy for the second hop:} The key part of the transmission strategy happens in the second hop. During the first time slot relay $v_2$ transmits $b_1$, relay $v_3$ transmits $b_2$, and relay $v_4$ transmits $b_3$ as depicted in Fig.~\ref{Fig:SecondHopEx1}. Ignoring the noise terms, relay $v_5$ obtains a linear combination of the symbols intended for destination $d_2$, $L_1\left( b_1,b_2,b_3 \right)$, that for simplicity we denote by $L_1(\vec{b})$. Similarly, relays $v_6$ and $v_7$ obtain linear combinations $L_2\left( b_1, b_2 \right)$ and $L_3\left( b_2, b_3 \right)$ respectively. During the first time slot, $v_1$ remains silent.

At this point, using the delayed knowledge of the channel state information, relay $v_3$ can (approximately) reconstruct $L_1(\vec{b})$. 
During the second time slot, relays $v_2$ and $v_4$ remain silent, relay $v_1$ sends out $a_1$, and relay $v_3$ sends out $L_1(\vec{b})$ (normalized to meet the power constraint). This way, $v_5$ obtains a linear combination of $a_1$ and $L_1(\vec{b})$ denoted by $L_4 (a_1, L_1(\vec{b}) )$. Note that $v_5$ already has access to $L_1(\vec{b})$, and thus it can recover $a_1$. Also, note that $v_6$ and $v_7$ obtain $L_1(\vec{b})$.

Finally, during the third time slot, relays $v_2$, $v_3$ and $v_4$ remain silent, and relay $v_1$ sends out $a_2$. Upon completion of these three time slots, $v_5$ has access to $a_1$ and $a_2$, $v_6$ has access to $L_1(\vec{b})$ and $L_2\left( b_1, b_2 \right)$, and $v_7$ has access to $L_1(\vec{b})$ and $L_3\left( b_2, b_3 \right)$.

\vspace{2mm}
\noindent {\bf Transmission strategy for the third hop and decoding:} The transmission strategy for the third hop is rather straightforward.
Relay $v_5$ sends $a_1$ and $a_2$ to $d_1$, and relays $v_6$ and $v_7$ send three linearly independent equations $L_1(\vec{b})$, $L_2\left( b_1, b_2 \right)$, and $L_3\left( b_2, b_3 \right)$ to $d_2$. Therefore $\left( D_1, D_2 \right) = \left( 2/3, 1 \right)$ DoF is achievable for the network of Fig.~\ref{Fig:3D1D2}.

\subsection{Example~2: A Two-Unicast Network with No Bottleneck Node}

\begin{figure}[ht]
\centering
\includegraphics[height = 2in]{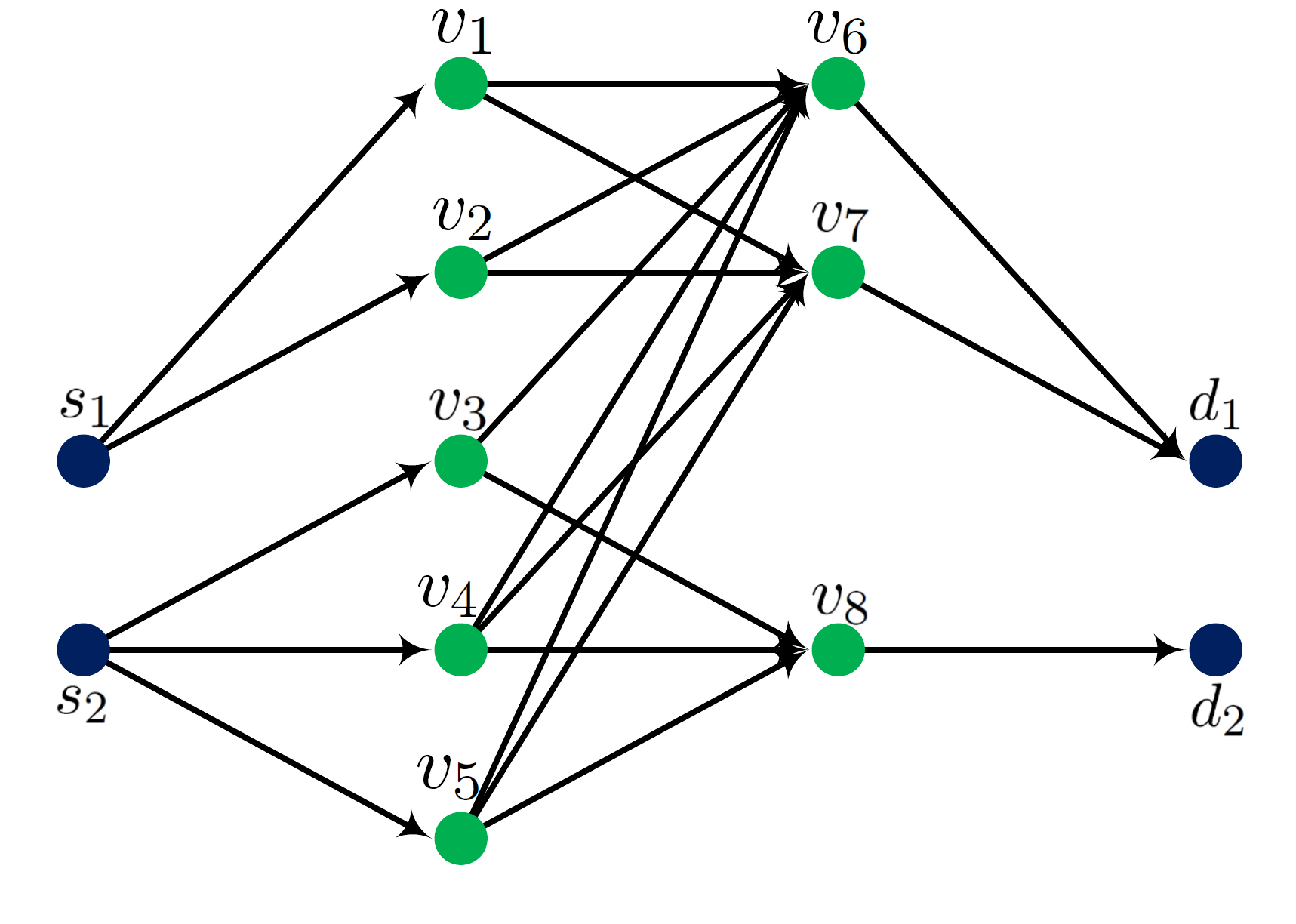}
\caption{In this example, we show we can achieve $\left( D_1, D_2 \right) = \left( 1, 1 \right)$.}\label{Fig:FullDoF}
\end{figure}

We now consider the network in Fig.~\ref{Fig:FullDoF}. As in the previous example, the lack of instantaneous CSIT prevents nodes $v_3$, $v_4$ and $v_5$ from scaling their signals according to the channel gains of the second hop so that their interference at $v_6$ and $v_7$ is canceled. Therefore, interference between the information flows is unavoidable. However, as we will show, since there is no single node acting as a bottleneck node (as in the previous example), $\left( 1, 1 \right)$ DoF can be achieved.  As it turns out, the diversity provided by an additional relay allows for a retroactive cancellation of the interference.

The transmission strategy has three time slots and the goal is for each source to communicate three symbols to its corresponding destination. 
For the first hop, the transmission strategy is very similar to that of the previous example and during each time slot, each source just sends a new symbol ($a_i$'s for source $s_1$ and $b_i$'s for source $s_2$, for $i=1,2,3$).

\begin{figure}[t]
\centering
\includegraphics[height=2.5in]{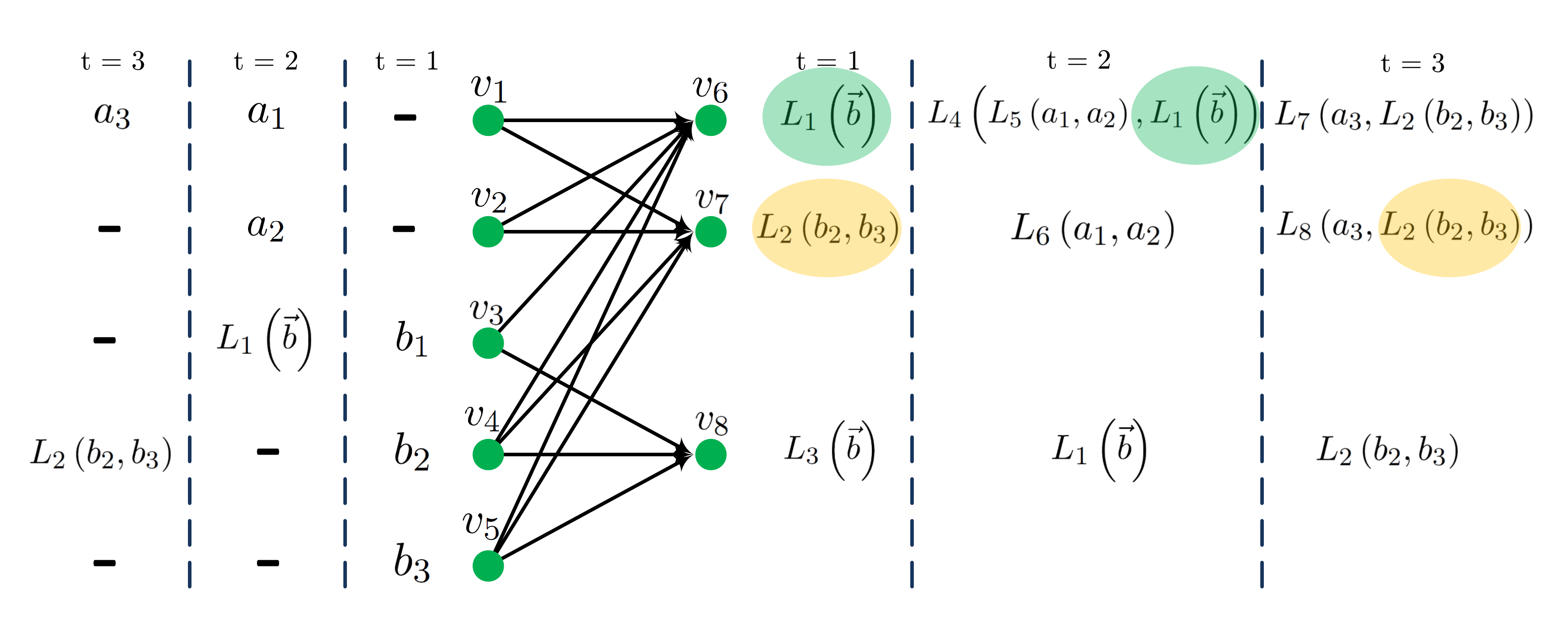}
\caption{Transmission strategy for the second hop of the network depicted in Fig.~\ref{Fig:FullDoF}.}\label{Fig:SecondHopEx2}
\end{figure}

\vspace{2mm}
\noindent {\bf Transmission strategy for the second hop:} Similar to the previous example, the key part of the transmission strategy is in the second hop and that is what we focus on. The transmission strategy is illustrated in Fig.~\ref{Fig:SecondHopEx2} and described below.

During the first time slot, relays $v_1$ and $v_2$ remain silent. Relay $v_3$ sends out $b_1$, relay $v_4$ sends out $b_2$, and relay $v_5$ sends out $b_3$. 
Ignoring the noise terms, relay $v_6$ obtains a linear combination of all symbols intended for destination $d_2$ that we denote by $L_1 ( \vec{b} )$. Similarly, relay $v_7$ obtains $L_2\left( b_2, b_3 \right)$ and relay $v_8$ obtains $L_{3}(\vec{b})$.

At this point, using the delayed knowledge of the channel state information, relay $v_3$ can reconstruct $L_1(\vec{b})$ and relay $v_4$ can reconstruct $L_2\left( b_2, b_3 \right)$. During the second time slot, $v_3$ sends out $L_1(\vec{b})$ and this equation becomes available to relay $v_8$. During this time slot, relay $v_1$ sends out $a_1$ and relay $v_2$ sends out $a_2$. Note that due to the connectivity of the network, relay $v_6$ receives $L_4\left(  L_5\left( a_1, a_2 \right), L_1(\vec{b}) \right)$, and relay $v_7$ receives $L_6\left( a_1, a_2 \right)$. Using the received signals during the first two time slots, relay $v_6$ can recover $L_5\left( a_1, a_2 \right)$. Relays $v_4$ and $v_5$ remain silent during the second time slot.

In the third time slot, relay $v_1$ sends out $a_3$, and relay $v_4$ sends out $L_2\left( b_2, b_3 \right)$. All other relays remain silent. This way, relays $v_6$ and $v_7$ obtain $L_{7}\left( a_3, L_2\left( b_2, b_3 \right) \right)$ and $L_{8}\left( a_3, L_2\left( b_2, b_3 \right) \right)$ respectively. Now note that using the received signal during time slots one and three, relay $v_7$ can recover $a_3$.

\vspace{2mm}
\noindent {\bf Transmission strategy for the third hop and decoding:} In the third hop, relays $v_6$ and $v_7$ can easily communicate $L_5\left( a_1, a_2 \right)$, $L_{6}\left( a_1, a_2 \right)$, and $a_3$ to destination $d_1$ during the three time slots. Note that these equations are (with probability one) linearly independent, and thus destination $d_1$ can recover its symbols. A similar story holds for destination $d_2$. This completes the achievability of $\left( D_1, D_2 \right) = \left( 1, 1 \right)$ for the network of Fig.~\ref{Fig:FullDoF}.


\section{Bottleneck Nodes}
\label{Section:Bottleneck}

As shown in the previous section, for the network of Fig.~\ref{Fig:FullDoF}, it is possible to exploit the diversity provided by the relays to retroactively cancel out the interference caused by relays $v_3$, $v_4$, and $v_5$ at relays $v_6$ and $v_7$. However, it is not difficult to see that the same approach cannot work for the network in Fig.~\ref{Fig:3D1D2}. This suggests that the network in Fig.~\ref{Fig:3D1D2} contains an informational bottleneck that is not present in the network in Fig.~\ref{Fig:FullDoF} and that restricts the sum DoF to be strictly less than $2$. 

\begin{figure}[ht]
\centering
\includegraphics[height = 4cm]{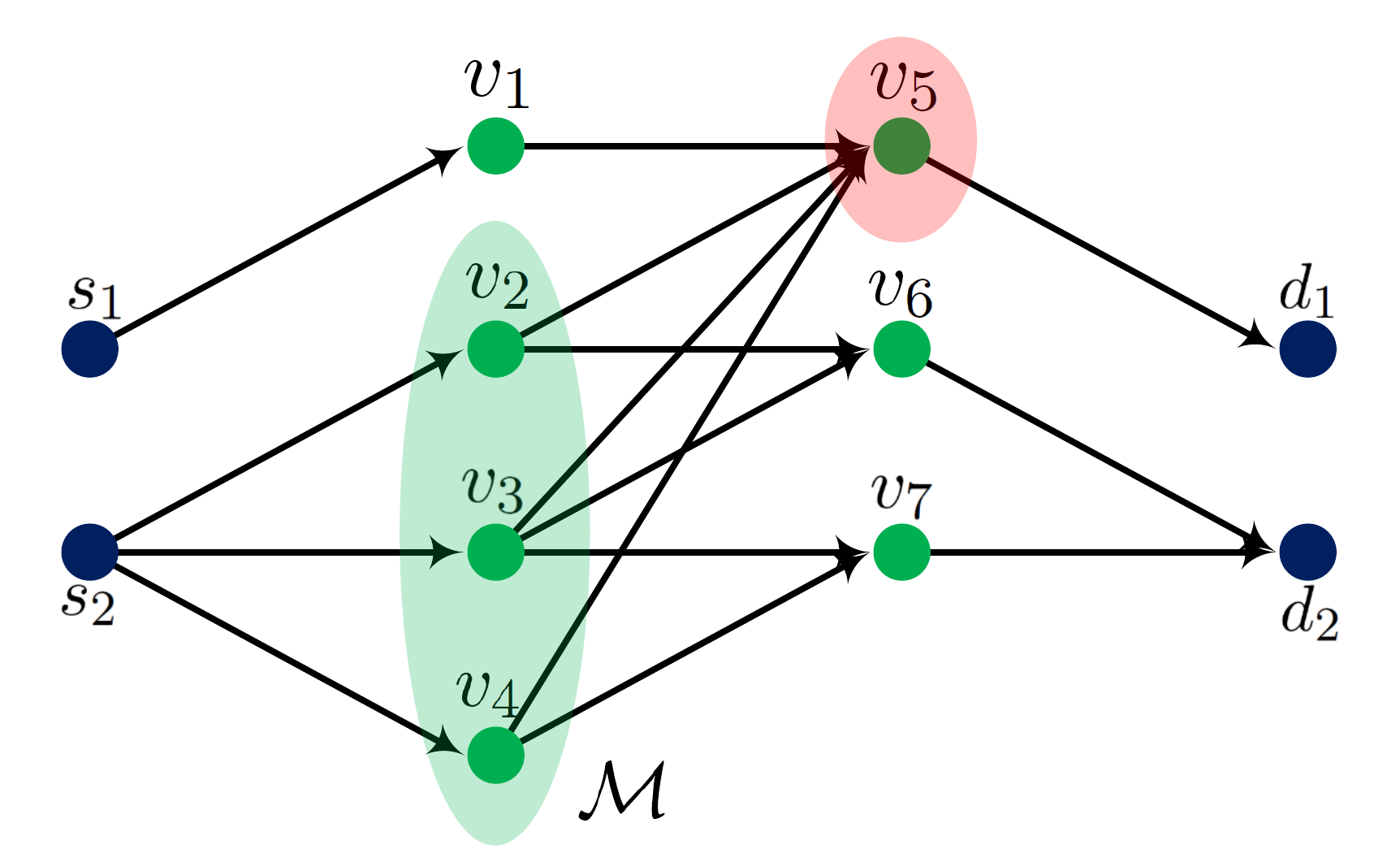}
\vspace{2mm}
\caption{$v_5$ acts as a bottleneck for the information flow. Relays $v_2$, $v_3$, and $v_4$ have to remain silent during a fraction of the time steps in order to allow $v_1$ and $v_5$ communicate. 
}
\label{Fig:3D1D2Bottleneck}
\end{figure}

As it turns out, this informational bottleneck is relay $v_5$. Notice that the information flow from $s_1$ to $d_1$ must go through $v_5$.
Moreover, the fact that the information flow from $s_2$ to $d_2$ must go through the set of nodes $\M = \{v_2,v_3,v_4\}$, and CSIT is obtained with delay, makes interference between the flows unavoidable and relays $v_2$, $v_3$, and $v_4$ have to remain silent during several time slots in order to allow $s_1$ and $d_1$ to communicate.  As we will show in this section, the \emph{rank} of transfer matrix between set $\M$ and the next layer determines how restrictive the bottleneck node $v_5$ is. For the example in Fig.~\ref{Fig:3D1D2Bottleneck}, since the rank is $3$, the bottleneck node implies a bound of the form $3 D_1 + D_2 \leq 3$.

Before stating the main result for bottleneck nodes, we need a few definitions.

\begin{definition}
A set of nodes $\A$, possibly a singleton, is a $(\B,\C)$-cut if the removal of $\A$ from the network disconnects all paths from $\B$ to $\C$.
\end{definition}

\begin{definition}
A node $v$ is an omniscient node if it is an $(\{s_1,s_2\},d_i)$-cut and there is a node $u \in \I(v) \cup \{v\}$ that is a $(s_{\bar i},\{d_1,d_2\})$-cut.
\end{definition}


The existence of an omniscient node imposes that the sum DoF is bounded by $1$, even when instantaneous CSIT is available. Motivated by the definition of an omniscient node, we introduce the notion of a $\rho$-bottleneck node, which reduces to an omniscient node for $\rho=1$.

\begin{definition} \label{transferdef}
For a set of nodes $\M$ in $\V_{\ell}$, $\ell = 1,2,\ldots,r-1$, let $F_{\M,\V_{\ell+1}}[t]$ be the transfer matrix between $\M$ and $\V_{\ell+1}$ at time $t$, $t=1,2,\ldots,n$.
\end{definition}

\begin{definition} \label{bottledef}
A node $v \in \V$ in layer $\ell+1$ is called a $\rho$-bottleneck node for $d_i$ if it is an $(\{s_1,s_2\},d_i)$-cut and there is a set $\M \subset \I(v)$ that is an $(s_{\bar i},\{d_1,d_2\})$-cut such that $\mathrm{rank}\left( F_{\M,\V_{\ell+1}}[t] \right) \overset{a.s.}= \rho$.
\end{definition}

We note that although a $1$-bottleneck node for $d_i$ is an omniscient node, the converse is not true. The following theorem provides an outer-bound on the DoF of a two-unicast network with delayed CSIT and a $\rho$-bottleneck node for $d_i$.

\begin{theorem}
\label{bottlethm}
Suppose a layered two-unicast wireless network $\N$ contains a $\rho$-bottleneck node for $d_i$, for $i \in \{1,2\}$. Then under the delayed CSIT assumption, we have 
\begin{align} \label{thmeq}
\rho D_{i} + D_{\bar{i}} \leq \rho.
\end{align}
\end{theorem}

Before providing the proof, we compare Theorem~\ref{bottlethm} to the outer-bounds in~\cite{vahid2015informational}, \emph{i.e.}
\begin{align} \label{eqsize}
\left| \M \right| D_{i} + D_{\bar{i}} \leq \left| \M \right|, \qquad i=1,2.
\end{align}

\begin{figure}[ht]
\centering
\subfigure[]{\includegraphics[height = 2in]{3D1D2.pdf}}
\hspace{.25in}
\subfigure[]{\includegraphics[height = 2in]{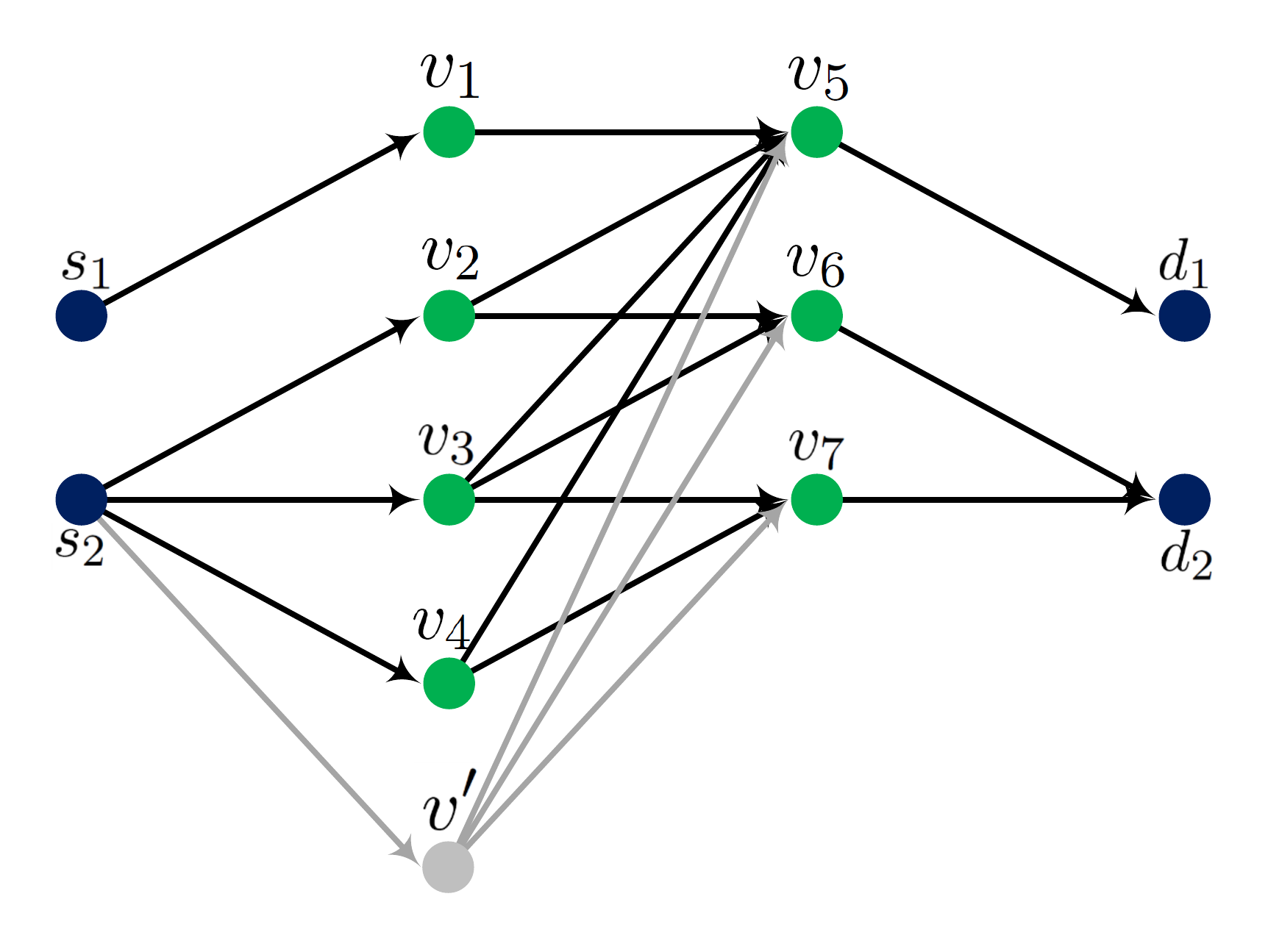}}
\caption{By adding a new node to the second layer of the network we studied in Example~1 of Section~\ref{Section:Examples}, we construct a network in which the results of~\cite{vahid2015informational} are loose. On the other hand, Theorem~\ref{bottlethm} provides tight outer-bounds.\label{Fig:setsizetorank}}
\end{figure}

\subsection{Comparison to Prior Results of~\cite{vahid2015informational}}
\label{Section:Compare}

The outer-bounds of Theorem~\ref{thmeq} are in general tighter than the ones given in (\ref{eqsize}) since
\begin{align}
\mathrm{rank}\left( F_{\M,\V_{\ell+1}}[t] \right) \overset{a.s.}= \rho \leq \left| \M \right|.
\end{align}

To understand how a bottleneck node affects the DoF region of two-unicast networks with delayed CSIT and to build intuition for our results, we revisit the network we studied in Example~1 of Section~\ref{Section:Examples}. This network is again depicted in Fig.~\ref{Fig:setsizetorank}(a). We construct a new network by adding a a new node to the second layer of this network as depicted in Fig.~\ref{Fig:setsizetorank}(b). We note that node $v_5$ is a $3$-bottleneck node for $d_1$ in both networks and $|\M| = 4$. The results of~\cite{vahid2015informational}, given in (\ref{eqsize}), imply that the DoF region is constrained by
\begin{align}
4 D_1 + D_2 \leq 4.
\end{align} 
However, Theorem~\ref{bottlethm} provides  
\begin{align}
3 D_1 + D_2 \leq 3,
\end{align} 
which is tighter. In fact, we can continue adding new nodes to the second layer of the network of Fig.~\ref{Fig:setsizetorank}(a) in a similar fashion and as the number of added nodes tends to infinity, the outer-bound in (\ref{eqsize}) gives us the trivial bound $D_1 \leq 1$. On the other hand, the results in Theorem~\ref{bottlethm} remain unchanged.

\subsection{Proof of Theorem~\ref{bottlethm}}

For $\rho=1$ the theorem follows since a $1$-bottleneck node for $d_1$ is an omniscient node.  In the remainder of this section, we prove this result for $\rho>1$.

\begin{figure}[ht]
\centering
\includegraphics[height = 4cm]{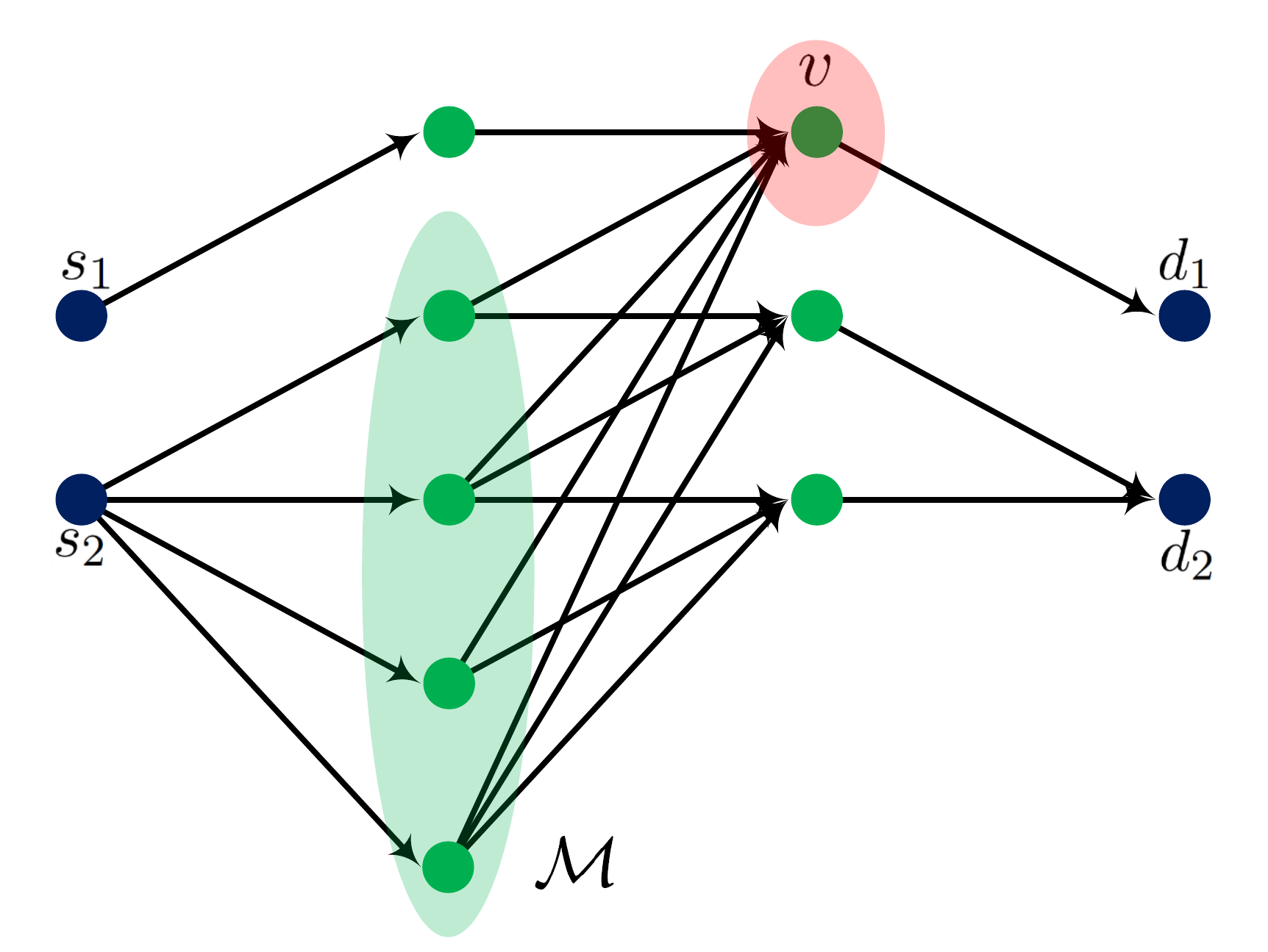}
\caption{A two-unicast network, $\N$, with a $3$-bottleneck node for destination $d_1$.}\label{Fig:3D1D2-Rank-nolabel}
\end{figure}

Suppose for network $\N$ we have a coding scheme that achieves $\left( D_1, D_2 \right)$ and $v$ is a $\rho$-bottleneck node for $d_1$ in layer $\V_{\ell+1}$. We use the network of Fig.~\ref{Fig:3D1D2-Rank-nolabel} to visualize our arguments. In this network it is straightforward to verify that node $v$ is a $3$-bottleneck node for $d_1$ according to Definition~\ref{bottledef}.

The proof contains two main steps stated in two separate lemmas. First, we construct a physically degraded multiple-input multiple-output (MIMO) broadcast channel (BC), $\Nmimo$, where it is possible to achieve any DoF pair $(D_1,D_2)$ that is achievable in the original network $\N$. Since the capacity of a physically degraded BC does not change with feedback, we can drop the delayed CSIT. The second step is then to show that when no CSIT is available, \eref{thmeq} must be satisfied in $\Nmimo$, which must therefore be satisfied in $\N$ as well. We now describe these two steps in more detail.

We first construct the MIMO BC $\Nmimo$ based on $\N$ as follows. The layer in $\N$ preceding the bottleneck node, $\V_\ell$, will become a single source $s'$ with $|\V_\ell|$ antennas. $\Nmimo$ will contain two receivers, namely $d_1^\prime$ and $d_2^\prime$. 

\begin{figure}[ht]
\centering
\includegraphics[height = 2in]{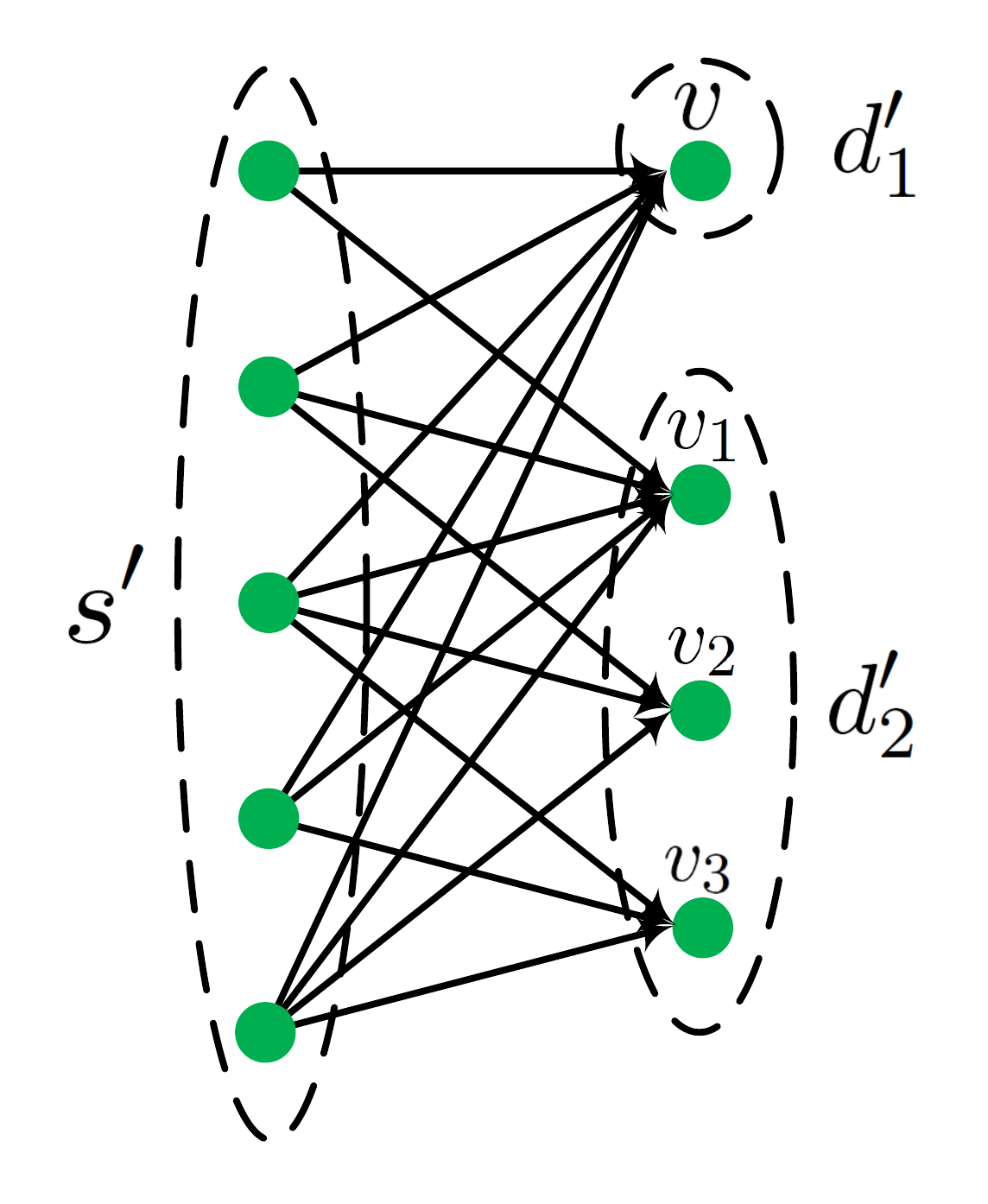}
\caption{Constructed MIMO BC $\Nmimo$ for the network $\N$ in Fig.~\ref{Fig:3D1D2-Rank-nolabel}.}\label{Fig:3D1D2BottleneckProof3}
\end{figure}

Receiver $d_1^\prime$ has only one antenna, which is a replica of the bottleneck node $v$ in $\N$. On the other hand, receiver $d_2^\prime$ has $\rho$ receive antennas labeled as $v_1,v_2,\ldots,v_\rho$. We note again that $\rho$ is the (almost sure) rank of the transfer matrix from $\M$ and layer $\ell+1$, \emph{i.e.} $\mathrm{rank}\left( F_{\M,\V_{\ell+1}}[t] \right) \overset{a.s.}= \rho$. For the remainder of this section and for simplicity, we drop the time index when no confusion is created.

Each row of $F_{\M,\V_{\ell+1}}$ determines the observed signal (minus the noise) of a node in layer $\ell+1$. Choose the row corresponding to node $v$ and $\rho-1$ other linearly independent rows of $F_{\M,\V_{\ell+1}}$. Denote the submatrix formed by these $\rho$ linearly independent row by $G_{\M,\V_{\ell+1}}$. The $\rho$ receive antennas of $d_2^\prime$ have the same connectivity, channel realizations and noise realizations as that of the nodes whose corresponding rows were chosen above. See Fig.~\ref{Fig:3D1D2BottleneckProof3} for a depiction. Without loss of generality, we assume the first receive antenna of $d_2^\prime$, $v_1$, has the same connectivity, channel realizations and noise realizations as that of node $v \in \N$. This guarantees that $Y_{v_1} = Y_v$, and that $\Nmimo$ is physically degraded. 

\begin{lemma} \label{bclem1}
Any DoF pair $(D_1,D_2)$ achievable in $\N$ is also achievable in $\Nmimo$.
\end{lemma}

\begin{proof}
First we focus on network $\N$, and assume we have a sequence of coding schemes that achieve a given rate pair $(R_1,R_2)$. Since node $v$ is a bottleneck node for $d_1$, it is an $(\{s_1,s_2\},d_1)$-cut and must be able to decode $W_1$ as well, and we have 
\begin{align}
\label{eq:d1cut}
H\left( W_1|Y_{v}^n, \H^n \right) \leq n \epsilon_n,
\end{align}
where $\ep_n \to 0$ as $n \to \infty$ from Fano's inequality. Next we notice that from the received signals in any given layer one should be able to reconstruct $W_2$, and we have 
\begin{align}
\label{eq:d2cut}
H\left( W_2| \left[ F_{\M,\V_{\ell+1}} X_{\M} + Z_{\V_{\ell+1}} \right]^n, X_{\M^c}^n, \H^n \right) \leq H \left( W_2| Y_{\V_{\ell+1}}^n, \H^n \right) \leq n \epsilon_n,
\end{align}
where as mentioned before $F_{\M,\V_{\ell+1}}$ is the transfer matrix between $\M$ and $\V_{\ell+1}$, and $\M^c = \V_{\ell} \setminus \M$. Our goal will be to emulate network $\N$ in the MIMO BC $\Nmimo$, so that destination $d_1'$ can recreate $Y_1^n$ to decode $W_1$, and destination $d_2'$ can approximately recreate $\left[ F_{\M,\V_{\ell+1}} X_{\M} + Z_{\V_{\ell+1}} \right]^n$ and $X_{\M^c}^n$ to decode $W_2$.

The main idea is to have source $s'$ in $\Nmimo$ simulate all the layers in $\N$ up to $\V_\ell$. In order to do that, let's first suppose that $s'$ and the destinations can share some randomness, drawn prior to the beginning of communication block. This shared randomness corresponds to noise and channel realizations for network $\N$ during a block of length $n$. Let us denote these noise and channel realizations by random vector $\U$. Notice that the channel and noise realizations in $\U$ are independent of the actual channel and noise realizations in $\Nmimo$. Using $\U$ and messages $W_1$ and $W_2$, $s'$ can transmit what the nodes in layer $\V_\ell$ from $\N$ would have transmitted (same distribution).

Since the received signal at $d_1'$ has the same distribution as the received signal at $v$ in network $\N$, similar to \eref{eq:d1cut}, for $\Nmimo$ we have 
\begin{align}
\label{eq:d1cutmimo}
H\left( W_1|Y_{d_1'}^n, \Hmimo^n, \U \right) \leq n \epsilon_n.
\end{align}

Moreover, since the first antenna of $d_2'$ receives the exact same signal as $d_1'$, we have
\begin{align}
H & \left( W_2 |Y_{d_2'}^n, \Hmimo^n, \U \right) \nonumber \\
& \leq H \left( W_2 |W_1, Y_{d_2'}^n, \Hmimo^n, \U \right) + H\left( W_1 |Y_{d_2'}^n, \Hmimo^n, \U \right) \nonumber \\
& \leq H \left( W_2 |W_1, Y_{d_2'}^n, \Hmimo^n, \U \right) + H\left( W_1 |Y_{d_1'}^n, \Hmimo^n, \U \right) \nonumber \\
& \overset{(\ref{eq:d1cutmimo})}\leq  H \left( W_2 |W_1, Y_{d_2'}^n, \Hmimo^n, \U \right) + n \epsilon_n. \label{eq:cut2}
\end{align}

Next, we notice that in $\N$, $X_{\M^c}$ is only a function of $\U$ and $W_1$. As a result, source $s'$ in $\Nmimo$ can reconstruct $X_{\M^c}$ and transmit it from the corresponding antenna in $\Nmimo$. This is because $\M$ is a $(s_{2},\{d_1,d_2\})$-cut in $\N$ and there can be no path from $s_2$ to $\M^c$.
Therefore, we have
\al{
& H \left( W_2|  W_1, Y_{d_2'}^n, \Hmimo^n, \U \right) \nonumber \\
& = H \left( W_2| \left[ F_{\M,\V_{\ell+1}} X_{\M} + \tilde Z_{\V_{\ell+1}} \right]^n, Y_{d_2'}^n, W_1, \Hmimo^n, \U \right) \nonumber \\
& \quad + I \left( W_2; \left[ F_{\M,\V_{\ell+1}} X_{\M} + \tilde Z_{\V_{\ell+1}} \right]^n | Y_{d_2'}^n, W_1, \Hmimo^n, \U \right) \nonumber \\
& = H \left( W_2| \left[ F_{\M,\V_{\ell+1}} X_{\M} + \tilde Z_{\V_{\ell+1}} \right]^n, X_{\M^c}^n, Y_{d_2'}^n, W_1, \Hmimo^n, \U \right) \nonumber \\
& \quad + I \left( W_2; \left[ F_{\M,\V_{\ell+1}} X_{\M} + \tilde Z_{\V_{\ell+1}} \right]^n | Y_{d_2'}^n, W_1, \Hmimo^n, \U \right) \nonumber \\
& \overset{(\ref{eq:d2cut})}\leq I \left( W_2; \left[ F_{\M,\V_{\ell+1}} X_{\M} + \tilde Z_{\V_{\ell+1}} \right]^n | Y_{d_2'}^n, W_1, \Hmimo^n, \U \right) + n \ep_n, \label{miterm}
}
where $\tilde{Z}_{\V_{\ell+1}}$ is a noise vector identically distributed as $Z_{\V_{\ell+1}}$ in $\N$ but independent from everything else. All we need to show is that the mutual information term in \eref{miterm} is $o(\log P)$.

Let $G_{\M,d_2'}$ and $G_{\M^c,d_2'}$ be the transfer matrices from $\M$ and $\M^c$ to $d_2'$ in $\Nmimo$. From $W_1$, $\Hmimo$, and $\U$, we can create
\begin{align} 
G_{\M^c,d_2'} X_{\M^c}, 
\end{align}
and use it to get
\aln{
G_{\M,d_2'} X_{\M} + \tilde{Z}_{\V_{\ell+1}} = Y_{d_2'} - G_{\M^c,d_2'} X_{\M^c}.
\label{Eq:RecoverXM}
}
Notice that $G_{\M,d_2'}$ is a $\rho \times |\M|$ matrix formed by choosing $\rho$ linearly independent rows of $F_{\M,\V_{\ell+1}}$. Moreover, recall that 
\begin{align}
\mathrm{rank}\left( F_{\M,\V_{\ell+1}} \right) \overset{a.s.}= \rho.
\end{align}
From (\ref{Eq:RecoverXM}) and the knowledge of channel realizations, we can compute
\begin{align}
F_{\M,\V_{\ell+1}} X_{\M} + \hat{Z}_{\V_{\ell+1}},
\end{align}
where $\hat Z$ is a combination of noise terms, whose power is a function of channel gains, but not of $P$.
Therefore, the mutual information term in \eref{miterm} can be upper bounded as 
\al{
& I \left( W_2; \left[ F_{\M,\V_{\ell+1}} X_{\M} + \tilde Z_{\V_{\ell+1}} \right]^n | Y_{d_2'}^n, W_1, \Hmimo^n, \U \right) \nonumber \\
& = I \left( W_2; \tilde Z_{\V_{\ell+1}}^n - \hat Z^n | Y_{d_2'}^n, W_1, \Hmimo^n, \U \right) \nonumber \\
& = h(\tilde Z_{\V_{\ell+1}}^n - \hat Z^n) - h\left( \tilde Z_{\V_{\ell+1}}^n \right) \leq n \, o(\log P), \label{ologpeq}
}
Therefore, from \eref{eq:cut2}, \eref{miterm} and \eref{ologpeq}, we have 
\begin{align}
H(W_2 | Y_{d_2'}^n, \Hmimo, \U) \leq n \ep_n + n \, o( \log P).
\end{align}
Hence, under the assumption of shared randomness, any pair $(D_1,D_2)$ achievable on $\N$ is also achievable in $\Nmimo$. But since the shared randomness is drawn independently from $W_1$ and $W_2$, we can simply fix a value $\U = {\bf u}$ for which the resulting error probability is at most the error probability averaged over $\U$. Thus the assumption of shared randomness can be dropped, and the lemma follows.
\end{proof}

Lemma~\ref{bclem1} allows us to bound the DoF of network $\N$ by instead bounding the DoF of $\Nmimo$.

\begin{lemma} \label{bclem2}
For the MIMO BC $\Nmimo$ defined above, we have
\begin{align}
\rho D_1 + D_2 \leq \rho.
\end{align}
\end{lemma}

\begin{proof}
The MIMO BC $\Nmimo$ is physically degraded since the first antenna of $d_2^\prime$ observes the same signal as $d_1^\prime$. We know that for a physically degraded broadcast channel, (Shannon) feedback does not enlarge the capacity region \cite{ElGamal-Degraded}. Therefore, we can ignore the delayed knowledge of the channel state information at the transmitter (\emph{i.e.} no CSIT assumption). We can further drop the correlation between the channel gains of the first receiver and the first antenna of the second receiver as the capacity of a BC only depends on the marginal distributions of the received signals~\cite{el2011network}. Thus for the MIMO BC described above under no CSIT, we have
\begin{align}
n & \left( \rho R_1 + R_2 - \epsilon_n \right) \nonumber \\
& \leq \rho I\left( W_1; Y_{d_1'}^n | \Hmimo^n\right) + I\left( W_2; Y_{d_2^\prime}^n | \Hmimo^n\right) \nonumber \\
& = \rho h\left( Y_{d_1'}^n | \Hmimo^n\right) - \rho h\left( Y_{d_1'}^n | W_1, \Hmimo^n\right) \nonumber \\
& \quad + h\left( Y_{d_2^\prime}^n | W_1, \Hmimo^n\right) - h\left( Y_{d_2^\prime}^n | W_1, W_2, \Hmimo^n\right) \nonumber \\
& \overset{(a)}= \rho h\left( Y_{d_1'}^n | \Hmimo^n\right)  - \rho h\left( Y_{d_1'}^n | W_1, \Hmimo^n\right) \nonumber \\
& \quad + h\left( Y_{d_2^\prime}^n | W_1, \Hmimo^n\right) - h\left( Z_{v_1}^n, \ldots, Z_{v_m}^n | \Hmimo^n\right) \nonumber \\
& \overset{(b)}= \rho h\left( Y_{d_1'}^n | \Hmimo ^n\right) - \rho h\left( Z_{d_1'}^n | \Hmimo^n\right)  \nonumber \\
& \quad + h\left( Y_{d_2^\prime}^n | W_1, \Hmimo^n\right) - \rho h\left( Y_{d_1'}^n | W_1, \Hmimo^n\right) \nonumber \\
& = \rho h\left( Y_{d_1'}^n | \Hmimo ^n\right) - \rho h\left( Z_{d_1'}^n | \Hmimo^n\right) + {h\left( Y_{v_{1}}^n | W_1, \Hmimo^n\right)}   \nonumber \\
& \quad + \ldots + h\left( Y_{v_{\rho}}^n | Y_{v_{1}}^n, \ldots, Y_{v_{\rho-1}}^n, W_1, \Hmimo^n\right) - \rho h\left( Y_{d_1'}^n | W_1, \Hmimo^n\right) \nonumber \\
& \overset{(c)}\leq \rho h\left( Y_{d_1'}^n | \Hmimo ^n\right) - \rho h\left( Z_{d_1'}^n | \Hmimo^n\right)  \nonumber \\
& \quad + \sum_{j=1}^{\rho}{ \left\{ h\left( Y_{v_{j}}^n | W_1, \Hmimo^n\right) - h\left( Y_{d_1'}^n | W_1, \Hmimo^n\right) \right\} } \nonumber \\
& \overset{(d)}\leq \rho \left\{ h\left( Y_{d_1'}^n | \Hmimo ^n\right) - h\left( Z_{d_1'}^n | \Hmimo^n\right) \right\} + n \rho o\left( \log P \right) \nonumber \\
& \leq \rho n \left( \tfrac12\log P + o\left( \log P \right) \right),
\end{align}
where $(a)$ follows since $X_{s'}^n$ is a function of $\left( W_1, W_2, \Hmimo^n \right)$; $(b)$ holds since noises are distributed as i.i.d. random variables; $(c)$ holds since conditioning reduces entropy; $(d)$ follows from Claim~\ref{Claim:AlignedImageSet} below. Dividing both sides by $n$ and taking the limit when $n \rightarrow \infty$, we get 
\begin{align}
\rho R_1 + R_2 \leq \rho \left( \tfrac12\log P + o\left( \log P \right) \right).
\end{align}
Therefore, from the discussion above and (\ref{Eq:DefinitionDoF}) we conclude that 
\begin{align}
\rho D_1+D_2 \leq \rho,
\end{align}
which completes the proof of Lemma~\ref{bclem2}.
\end{proof}

\begin{claim}
\label{Claim:AlignedImageSet}
For the MIMO BC $\Nmimo$ defined above with no CSIT, we have
\begin{align}
h\left( Y_{v_j}^n | W_1, \Hmimo^n\right) - h\left( Y_{d_1'}^n | W_1, \Hmimo^n\right) \leq n o\left( \log P \right), \qquad j=1,2,\ldots,\rho.
\end{align}
We note that $v_1,v_2,\ldots,v_\rho$ in $\Nmimo$ are the $\rho$ receive antennas of $d_2'$.
\end{claim}

\begin{figure}[ht]
\centering
\includegraphics[height = 2.5in]{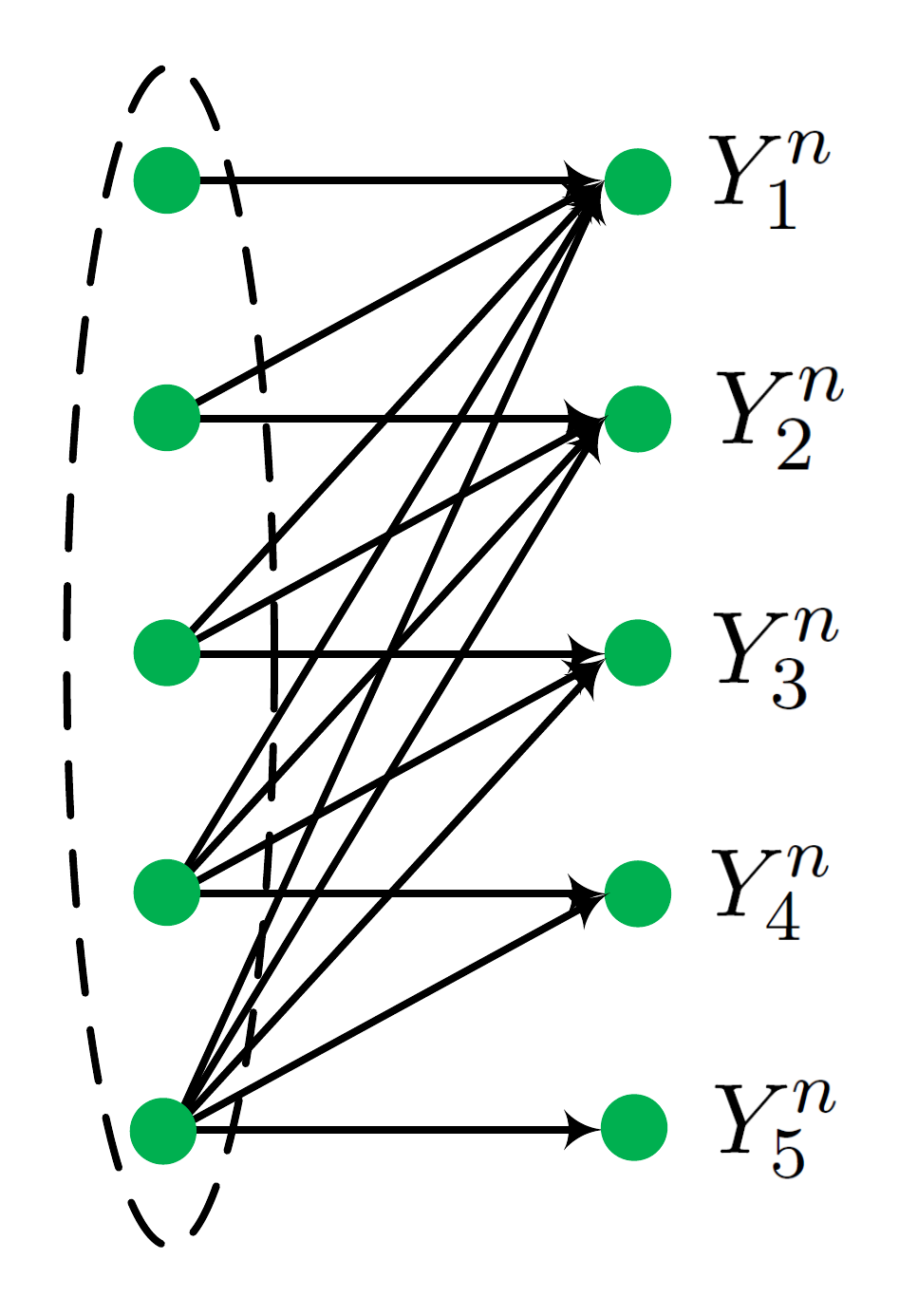}
\caption{Consider the MIMO BC of Fig.~\ref{Fig:3D1D2BottleneckProof3}. We construct a BC with the same number of transmit antennas, $5$ in this case, and with $5$ single-antenna receivers. Under the no CSIT assumption and for the same input distribution, $Y_1^n$ is statistically the same as $Y_{d_1'}^n$. Moreover, for any antenna in $d_2'$ of the MIMO BC of Fig.~\ref{Fig:3D1D2BottleneckProof3}, there is a counterpart in this network. For instance, $Y_3^n$ is statistically the same as $Y_{v_2}^n$.}\label{Fig:BC-Jafar}
\end{figure}

\begin{proof}
The proof of this claim follows from the results of~\cite{davoodi2016aligned} as described below. Fix the MIMO BC $\Nmimo$ with $|\V_{\ell}|$ antennas and $\rho$ single-antenna receivers where $|\V_{\ell}|$ and $\rho$ are derived from the underlying two-unicast network $\N$. Consider the broadcast channel depicted in Fig.~\ref{Fig:BC-Jafar} with $|\V_{\ell}|$ transmit antennas and the same number of single-antenna receivers with partial connectivity (transmit antenna $j$ is connected to receiver $1,2,\ldots,j$). The channel gains at each time are real-valued i.i.d. random processes obeying the distribution as the channel gains in $\N$. For this network under the no CSIT assumption, the authors in~\cite{davoodi2016aligned} prove that (see (57 in~\cite{davoodi2016aligned}))\footnote{We slightly abuse the notation and use $\Hmimo^n$ both for this network and for $\Nmimo$.}:
\begin{align}
\label{Eq:DavoodiIneq}
h\left( Y_{j}^n | W_1, \Hmimo^n\right) \geq h\left( Y_{j+1}^n | W_1, \Hmimo^n\right) + n o\left( \log P \right), \qquad j=1,2,\ldots,\rho-1.
\end{align}

For the same input distribution in $\Nmimo$ and the BC constructed above\footnote{A trivial relabeling of indices might be needed for the transmit antennas.}, we have:
\begin{align}
& 1)~h\left( Y_{d_1'}^n | W_1, \Hmimo^n\right) = h\left( Y_{1}^n | W_1, \Hmimo^n\right); \nonumber \\
& 2)~\forall~v_i \in \Nmimo,~\exists~j \in \{1,\ldots,\rho\}:~h\left( Y_{v_i}^n | W_1, \Hmimo^n\right)~=~h\left( Y_{j}^n | W_1, \Hmimo^n\right).
\end{align}
Using this observation and (\ref{Eq:DavoodiIneq}), the proof of the claim follows immediately.
\end{proof}

Lemma~\ref{bclem1} and Lemma~\ref{bclem2} complete the proof of Theorem~\ref{bottlethm}.

\begin{figure}[ht]
\centering
\includegraphics[height = 4cm]{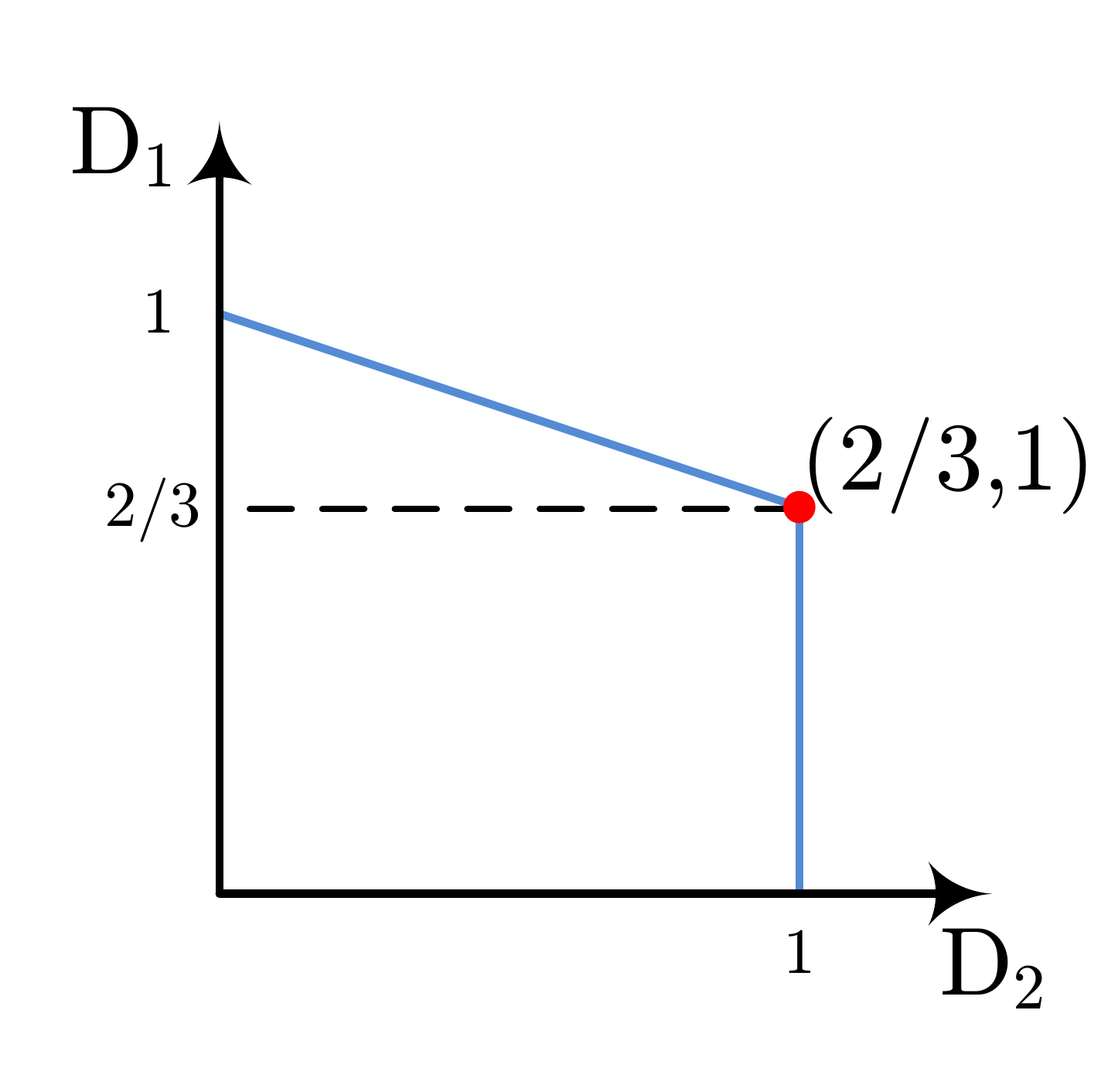}
\caption{The degrees-of-freedom region of the layered two-unicast network of Fig.~\ref{Fig:3D1D2} with delayed CSIT.}\label{Fig:Region3D1D2}
\end{figure}

Now consider again the network of Fig.~\ref{Fig:3D1D2}.  In this network, $v_5$ is a $3$-bottleneck node for $d_1$. Thus, for this network using Theorem~\ref{bottlethm}, we have
\begin{equation}
\label{eq:DoF3D1D2}
\left\{ \begin{array}{ll}
\vspace{1mm} 0 \leq D_i \leq 1, & i = 1,2, \\
3 D_1 + D_2 \leq 3. &
\end{array} \right.
\end{equation}
This region is depicted in Fig.~\ref{Fig:Region3D1D2}. In Section~\ref{Section:Examples}, we provided the achievability proof of corner point $\left( D_1, D_2 \right) = \left( 2/3, 1 \right)$. As a result, the outer-bound provided by Theorem~\ref{bottlethm} (alongside individual bounds) completely characterizes the achievable DoF region in this case.

\section{Proof of Theorem~\ref{mainthm}}
\label{Section:Proof}

In this section we describe the proof of Theorem~\ref{mainthm}. In essence, we show that the example considered in Section~\ref{bottleexsec} can be generalized to a class of networks that contain bottleneck nodes whose corresponding outer bounds can be achieved.

The proof has two steps: 1) we construct a network in which a bottleneck node for $d_1$ exists, and we show that outer-bound 
\begin{align}
\label{Eq:Step1}
\rho D_1 + D_2 \leq \rho,
\end{align}
is tight; 2) we concatenate this network with a similar network in which the indices are flipped, and we show that in this concatenate network both (\ref{Eq:Step1}) and 
\begin{align}
\label{Eq:Step2}
D_1 + \rho D_2 \leq \rho,
\end{align}
are tight, and thus completing the proof.

\begin{figure}[ht]
\centering
\includegraphics[height = 2in]{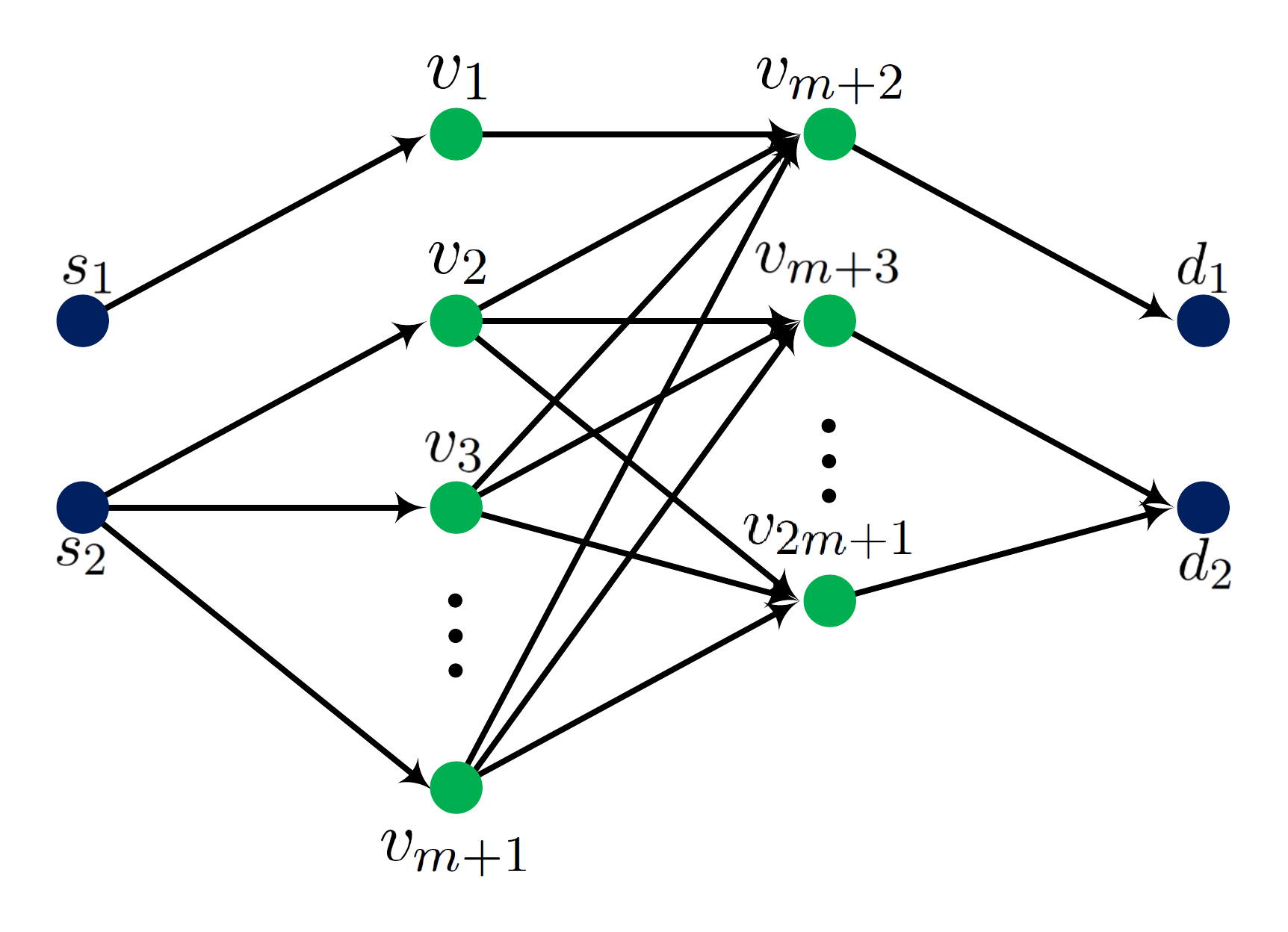}
\caption{In this example we show that we can achieve corner point $\left( D_1, D_2 \right) = \left( (m-1)/m, 1\right)$.}\label{Fig:mD1D2}
\end{figure}

\vspace{2mm}
\noindent {\bf Step~1}: Consider the network illustrated in Fig.~\ref{Fig:mD1D2}. In this network wireless nodes are organized in four layers:
\begin{align}
& \V_1 = \{ s_1, s_2 \}, \qquad \V_2 = \{ v_1,v_2, \ldots,v_{m+1} \}, \nonumber \\
& \V_3 = \{ v_{m+2}, \ldots,v_{2m+1} \}, \qquad \V_4 = \{ d_1, d_2 \}.
\end{align}
In the second layer $v_1$ is only connected to $v_{m+2}$, and any other node  in $\V_2$ is connected to \emph{all} nodes in $\V_3$. Based on Definition~\ref{bottledef}, $v_{m+2}$ is an $m$-bottleneck node for $d_1$ with
\begin{align}
\M = \{ v_2, v_3, \ldots,v_{m+1} \}.
\end{align}
Moreover, $F_{\M,\V_3}$ is a full-rank $m \times m$ square matrix. We note that there is no bottleneck node for $d_2$ in the network of Fig.~\ref{Fig:mD1D2}. 

We show that for this network we can achieve corner point 
\begin{align}
\left( D_1, D_2 \right) = \left( \frac{m-1}{m}, 1\right).
\end{align}
The achievability strategy is a generalization of the strategy presented for the network of Fig.~\ref{Fig:3D1D2}, and uses $m$ time steps. As in that case, the transmission scheme for the first and third hops is straightforward and we only focus on the second hop.

\noindent {\bf Transmission strategy for the intermediate problem:} The transmission strategy has $m$ time slots. During the first time slot, relay $v_1$ remains silent and relay $v_j$ sends out symbol $b_{j-1}$ intended for destination $d_2$, $j=2,3,\ldots,m+1$. Ignoring the noise terms, relay $v_j$ obtains a linear combination of the symbols intended for destination $d_2$, $L_{j-m-1}\left( \vec{b} \right)$, $j = m+2, \ldots,2m+1$.

\begin{figure}[ht]
\centering
\includegraphics[height = 2in]{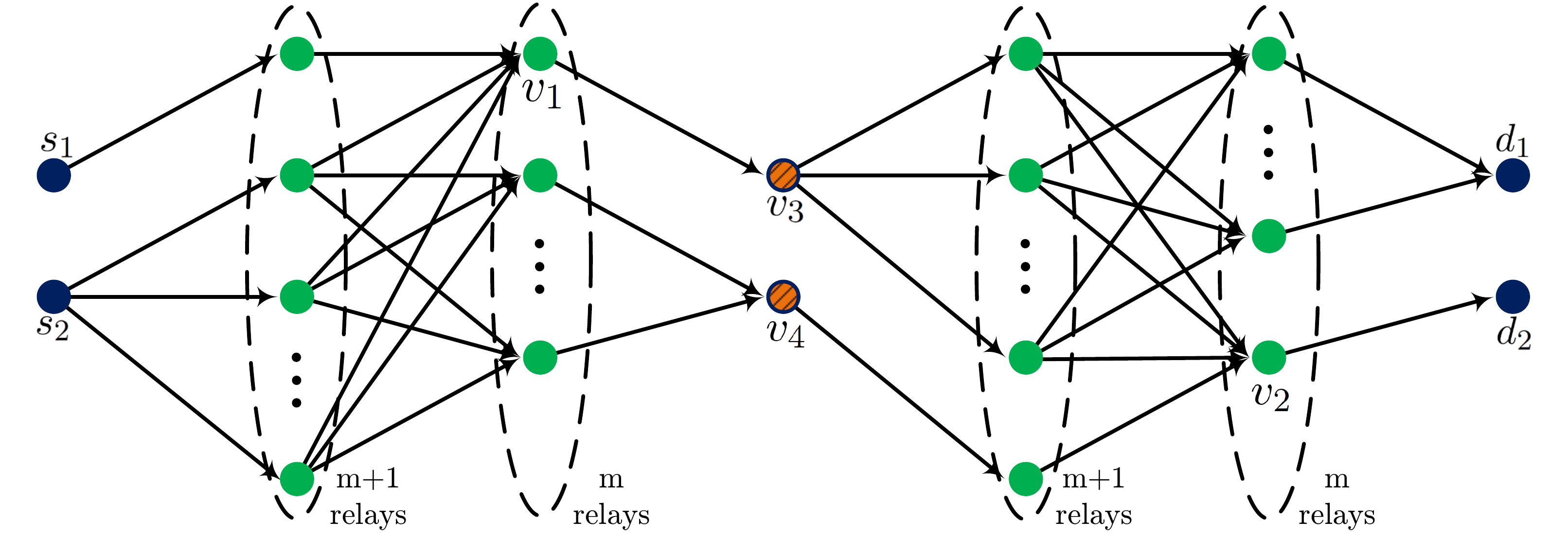}
\caption{Relay $v_1$ is an $m$-bottleneck node for $d_1$ and relay $v_2$ is an $m$-bottleneck node for $d_2$.} \label{Fig:TwoBounds}
\end{figure}

At this point, using the delayed knowledge of the channel state information, relay $v_2$ can (approximately) reconstruct $L_1(\vec{b})$. During the second time slot, relay $v_1$ sends out $a_1$, relay $v_3$ sends out $L_1(\vec{b})$ (normalized to meet the power constraint), and relays $v_3,\ldots, v_{m+1}$ remain silent. This way, $v_{m+2}$ obtains a linear combination of $a_1$ and $L_1(\vec{b})$ denoted by $L_{m+1} (a_1, L_1(\vec{b}) )$. Note that $v_{m+2}$ already has access to $L_1(\vec{b})$ and thus can recover $a_1$. Also, $L_1(\vec{b})$ becomes available to $v_j$ for $j=2,3,\ldots,m+1$.

During time slot $\ell$, $\ell = 3, \ldots, m$, relay $v_1$ sends out $a_{\ell-1}$ and relays $v_2, v_3, \ldots, v_{m+1}$ remain silent. Note that with this strategy, $v_{m+2}$ obtains $a_1, a_2, \ldots, a_{m-1}$, and relays $v_2, v_3, \ldots, v_{m+1}$ (with probability 1) obtain $m$ linearly independent combinations of $b_1, b_2, \ldots, b_m$. Then the task for the third hop is to simply deliver $a_1, ..., a_{m-1}$ to $d_1$ and the $m$ linearly independent combinations of $b_1, ..., b_m$ to $d_2$.


Since we have matching inner and outer bounds, we conclude that for the network in Fig.~\ref{Fig:mD1D2}, the sum DoF are $\dE = 1 + (m-1)/m = 2 - 1/m$, for $m \in \{1,2,...\}$.
Notice that this corresponds to half of the values in the set $\S$ in \eref{Seq}.
To obtain the remaining values in $\S$, we need a class of networks that contain both a bottleneck node for $d_1$ and a bottleneck node for $d_2$.

\begin{figure}[ht]
\centering
\includegraphics[height = 2in]{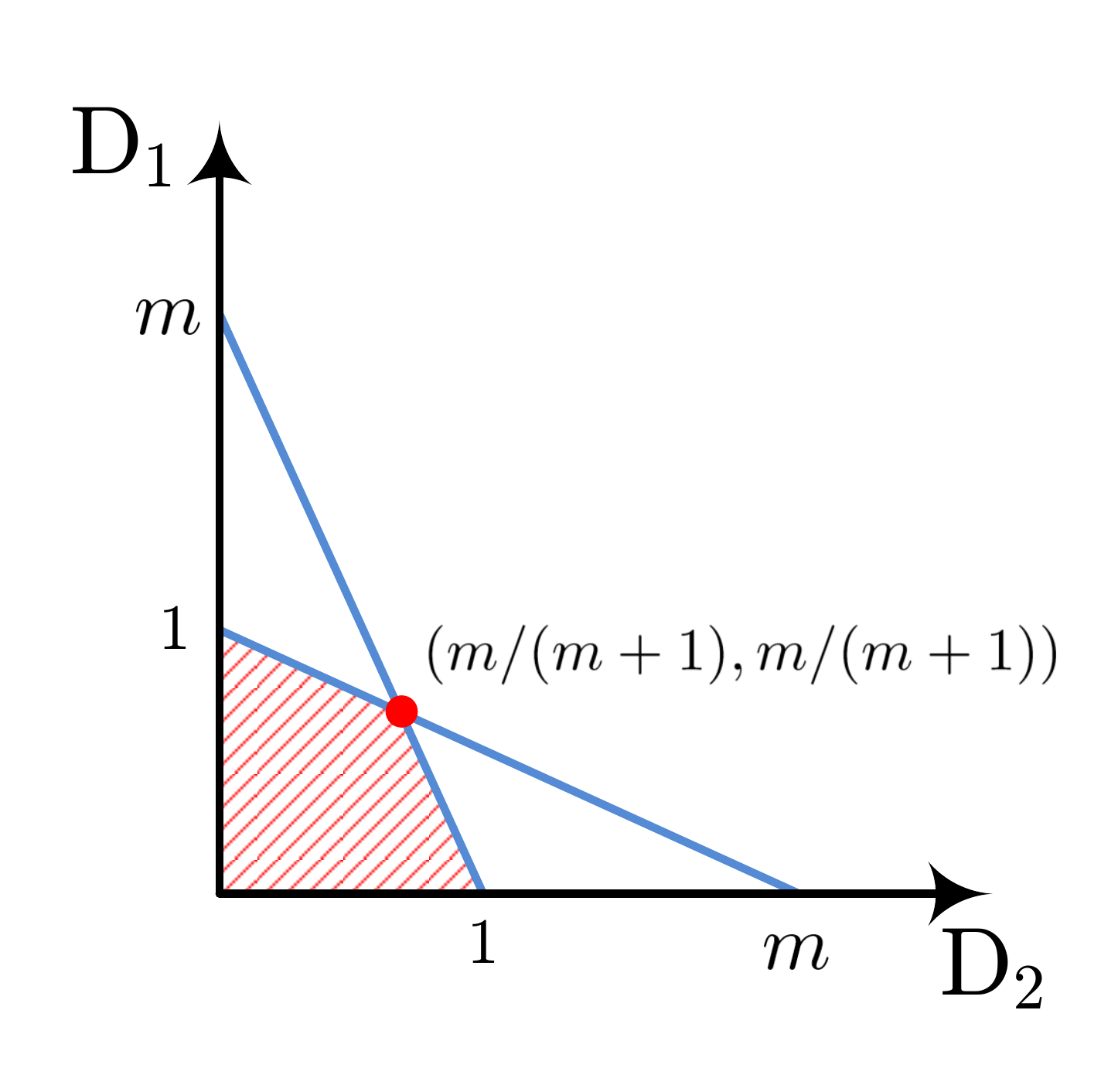}
\caption{The region described by the outer-bounds in (\ref{Eq:ProofTwoBounds}).} \label{Fig:TwoBoundsRegion}
\end{figure}

\vspace{2mm}
\noindent {\bf Step~2}: Consider the network depicted in Fig.~\ref{Fig:TwoBounds}. For simplicity of notation, we have only labeled a few relays in this network. 
We claim that for this network $\dE = 2m/(m+1)$, $m \in \{1,2,...\}$.
First, we prove the converse. It is straightforward to verify that relay $v_1$ is an $m$-bottleneck node for $d_1$ and relay $v_2$ is an $m$-bottleneck node for $d_2$. Thus from Theorem~\ref{bottlethm}, we have 
\al{ \label{Eq:ProofTwoBounds}
m D_{i} + D_{\bar{i}} \leq m, \qquad i=1,2.
}
The region described by these two outer-bounds is depicted in Fig.~\ref{Fig:TwoBoundsRegion}. To prove that the outer-bounds are tight, it suffices to prove the achievability of corner point 
\begin{align}
\left( D_1, D_2 \right) = \left( m/(m+1), m/(m+1) \right).
\end{align} 

\vspace{2mm}
\noindent {\bf Transmission strategy:} The goal is to deliver $m$ symbols to each destination during $m+1$ time slots. Denote the symbols intended for $d_1$ by $a_i$'s and the symbols intended for $d_2$ by $b_i$'s, $i=1,2,\ldots,m$. 
We point out that the network in Fig.~\ref{Fig:TwoBounds} can be seen as a concatenation of the network in Fig.~\ref{Fig:mD1D2} with flipped copy of itself.
Hence, we will describe the achievability in terms of each of the two subnetworks.
We first describe how to deliver $a_i$'s to relay $v_3$ and $b_i$'s to relay $v_4$. 
Then, the goal becomes for relay $v_3$ to deliver $a_i$'s to $d_1$ and for relay $v_4$ to deliver $b_i$'s to $d_2$, $i=1,2,\ldots,m$. 
Since the two subnetworks are essentially identical, 
we only need to show that we can deliver $a_i$'s to relay $v_3$ and $b_i$'s to relay $v_4$ during $m+1$ time slots. 
Then, the relays in the second subnetwork 
will implement a similar strategy to that of the nodes in the first subnetwork. 

Since the first subnetwork is identical to the network of Fig.~\ref{Fig:mD1D2}, by using the same strategy, during $m$ time slots we can deliver $m-1$ symbols to $v_3$ and $m$ symbols to $v_4$. 
During the last time slot, \emph{i.e.} time slot $m+1$, source $s_2$ remains silent, and source $s_1$ sends out one more symbol, $a_m$, to relay $v_3$. 
This way, we successfully deliver $a_i$'s to relay $v_3$ and $b_i$'s to relay $v_4$
during $m+1$ time slots, $i=1,2,\ldots,m$. 
Repeating the same strategy over the second subnetwork, each destination can decode its $m$ symbols over $m+1$ time steps, and we conclude 
that $\dE = 2m/(m+1) = 2-2/(m+1)$.
This completes the proof of Theorem~\ref{mainthm}. 

\section{Discussion}
\label{Section:Discussion}

In this paper we introduced a new technique to derive outer bounds on the DoF of two-unicast wireless networks with delayed CSIT, and we presented several transmission strategies that can achieve these outer bounds.
The presented transmission strategies achieve the optimal DoF in a finite number of time slots. In this section, we discuss two follow-up questions to our main results:
\begin{enumerate}
\item Do bounds of the form $m D_i + D_{\bar{i}} \leq m$ for $m \geq 1$ suffice to characterize the DoF region of the two-unicast wireless networks with delayed CSIT?
\item Can we achieve the optimal DoF region of a two-unicast wireless networks with delayed CSIT in a finite and bounded number of time slots?
\end{enumerate}

\begin{figure}[ht]
\centering
\includegraphics[height = 3cm]{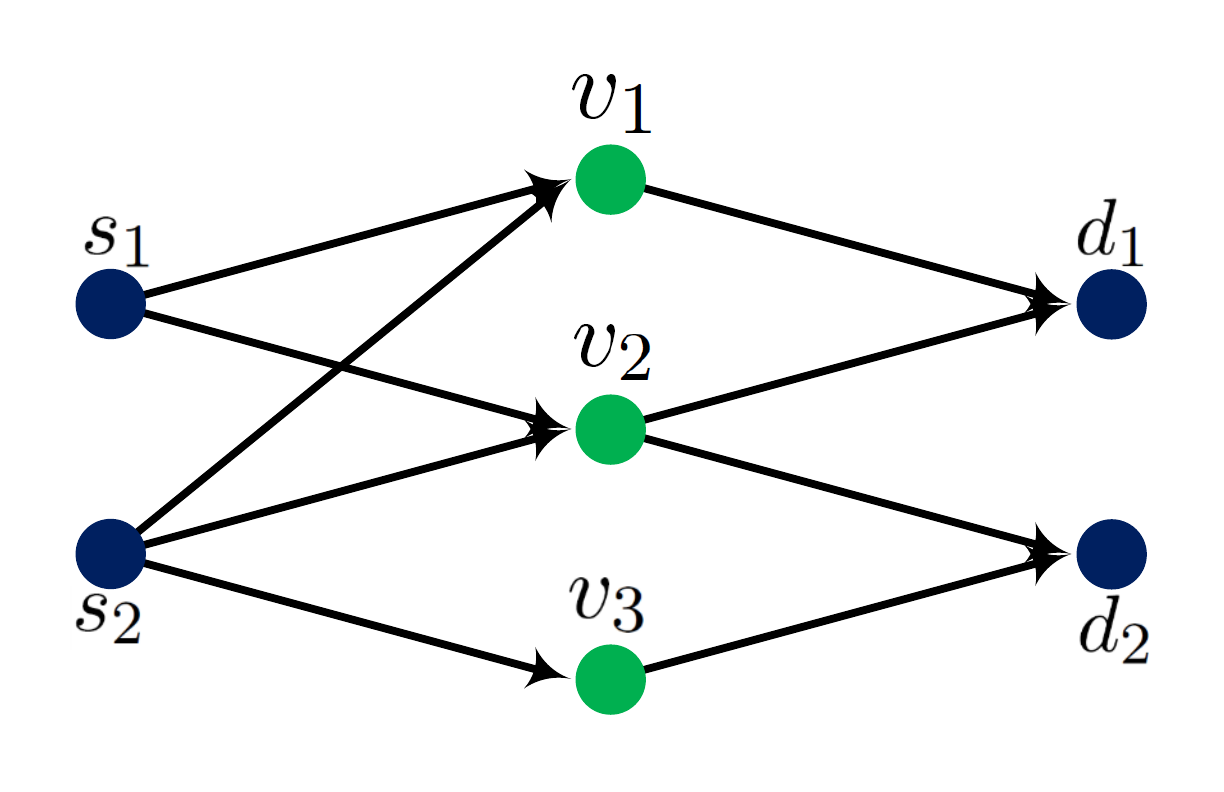}
\caption{The outer-bound provided by Theorem~\ref{bottlethm} does not describe the DoF region of this network. Moreover, the achievability strategy for the corner points of the DoF region does not have a finite number of time slots.}\label{Fig:D1D2onehalf}
\end{figure}

As it turns out, the answers to the questions posed above are both negative. To provide some insights, we consider the network depicted in Fig.~\ref{Fig:D1D2onehalf}. Under instantaneous CSIT assumption, the DoF region of this network is derived in~\cite{dof2unicastfull} and is given by
\begin{equation}
\label{eq:DoFRegionIlan}
\left\{ \begin{array}{ll}
\vspace{1mm} 0 \leq D_i \leq 1, & i = 1,2, \\
D_1 + D_2 \leq \frac{3}{2}. &
\end{array} \right.
\end{equation}
Interestingly, under the delayed CSIT assumption, we can still achieve this region. 
However, the network in Fig.~\ref{Fig:D1D2onehalf} contains no bottleneck nodes.
Moreover, it can be verified that the region in \eref{eq:DoFRegionIlan} cannot be obtained 
from bounds of the form $m D_i + D_{\bar{i}} \leq m$ for $m \geq 1$. 

Next, we briefly describe the achievability strategy for corner point $\left( D_1, D_2 \right) = \left( 1, 1/2 \right)$. The achievability strategy goes over $2m+1$ time slots and upon completion of the transmission, we achieve
\begin{align}
\left( D_1, D_2 \right) = \left( \frac{2m}{2m+1}, \frac{k}{2m+1} \right),
\end{align}
where $m$ is an arbitrarily chosen parameter.
Thus, as the number of time slots $m$ goes to infinity, we achieve arbitrarily close to the corner point $\left( D_1, D_2 \right) = \left( 1, 0.5 \right)$.

\begin{figure*}[t]
\centering
\subfigure[]{\includegraphics[width = \columnwidth]{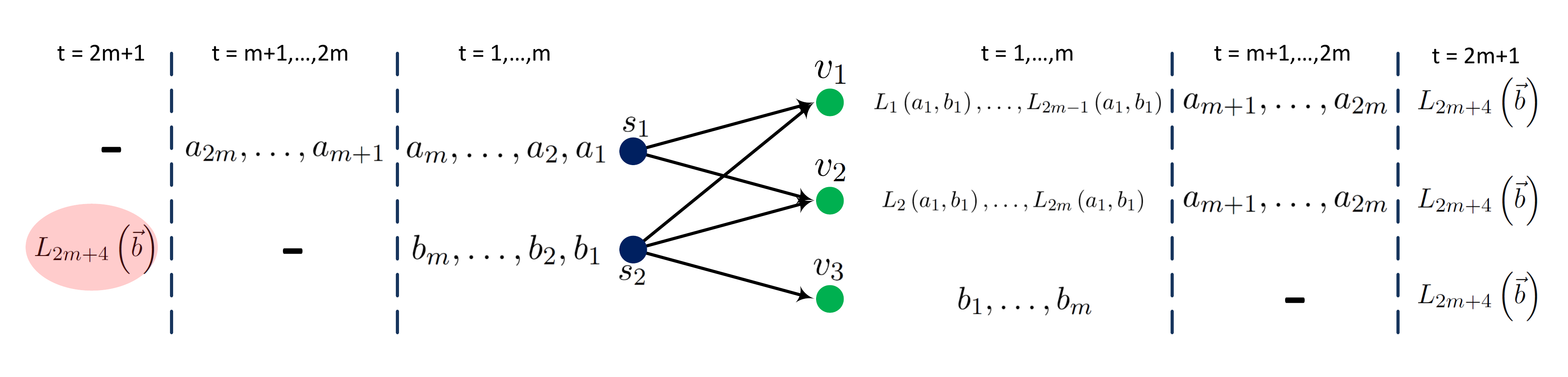}}
\subfigure[]{\includegraphics[width = \columnwidth]{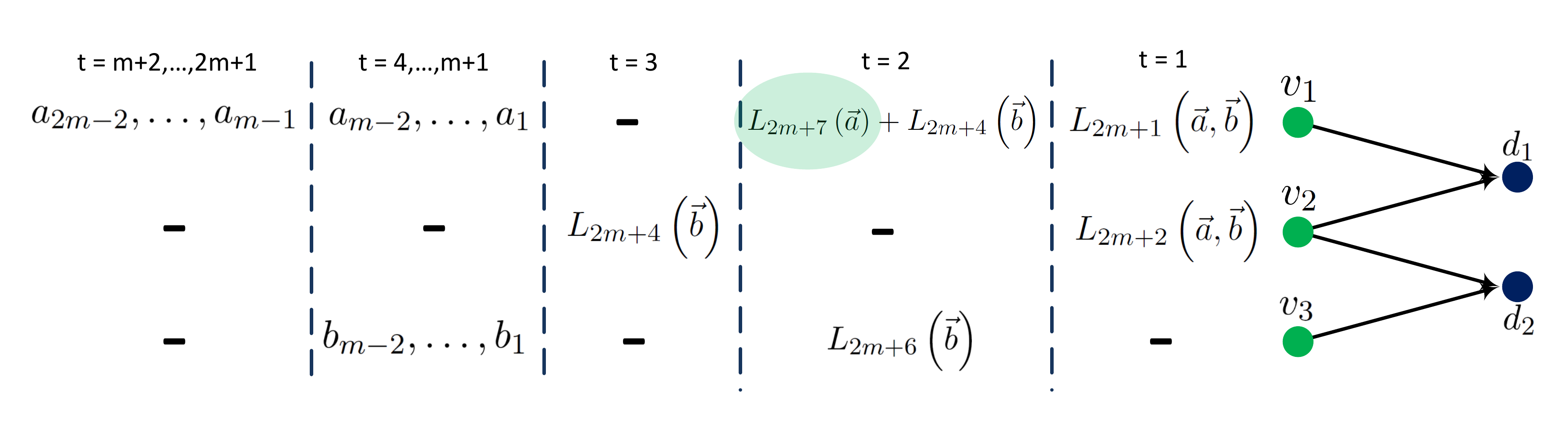}}
\subfigure[]{\includegraphics[width = \columnwidth]{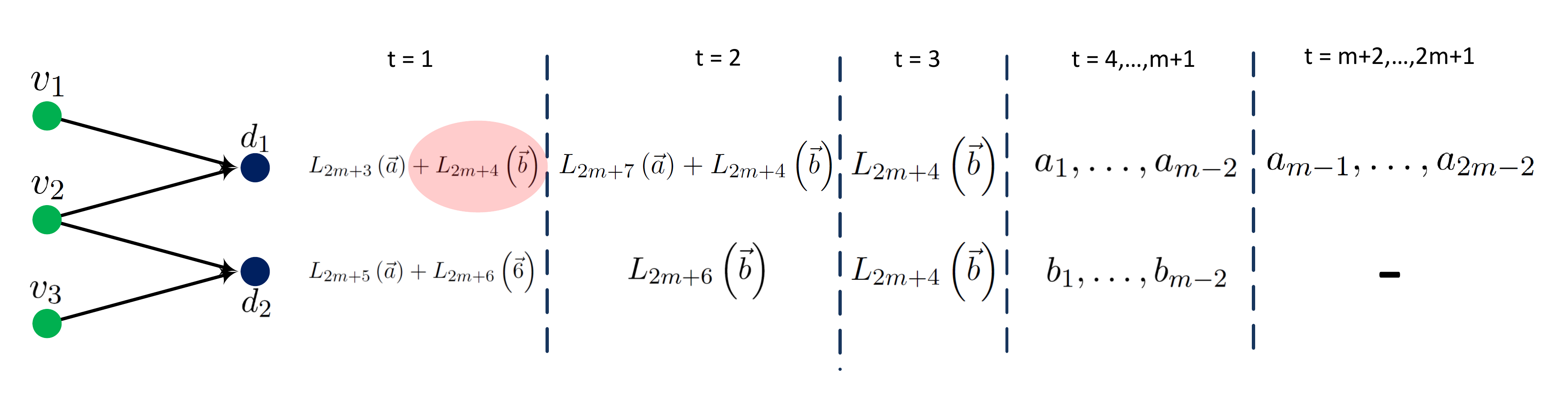}}
\caption{Achievability strategy for corner point $\left( D_1, D_2 \right) = \left( 1, 1/2 \right)$ of the DoF region of the network of Fig.~\ref{Fig:D1D2onehalf}: (a) strategy for the first hop; (b) transmit signal for the second hop; (c) receive signal for the second hop. The achievability strategy uses $2m+1$ time slots and as $m \to \infty$, we achieve the desired corner point.} \label{Fig:D1D2onehalf-Cornerpoint}
\end{figure*}

The transmission strategy is illustrated in Fig.~\ref{Fig:D1D2onehalf-Cornerpoint}. We highlight the important aspects of this strategy here. First we note that by interleaving different blocks, we encode such that the first $2m$ time slots of the first hop occur before the first time slot of the second hop. This way, there will be no issues regarding causality in the network.

For the first hop, the communication during the first $2m$ time slots is straightforward. In the second hop during the first time slot, relays $v_1$ and $v_2$ create random linear combinations of all the signals they received during the first $2m$ time slots of the first hop and send them out. Destination one obtains $L_{2m+3}\left(\vec{a}\right) + L_{2m+4}(\vec{b})$, and destination two obtains $L_{2m+5}\left(\vec{a}\right) + L_{2m+6}(\vec{b})$. Our goal is to deliver $L_{2m+4}(\vec{b})$ to both receivers. Relay $v_3$ can reconstruct $L_{2m+4}(\vec{b})$, however, there is no link form $v_3$ to destination one. As a result, during the final time slot, the second source sends out $L_{2m+4}(\vec{b})$ and this signal becomes available to all receivers (see Fig.~\ref{Fig:D1D2onehalf-Cornerpoint} where $L_{2m+4}(\vec{b})$ is highlighted by a red oval). 

The key idea for the achievability would be the observation that relay $v_1$ can combine its previous observations in a way that $b_i$'s form $L_{2m+4}(\vec{b})$, $i=1,2,\ldots,m$. This way, during the first two time slots, the interference at destination one would be the same. Thus, if we provide $L_{2m+4}(\vec{b})$ to destination one, it can recover $L_{2m+3}\left(\vec{a}\right)$ and $L_{2m+7}\left(\vec{a}\right)$. Finally, we note that $L_{2m+6}(\vec{b})$ is linear combination of $b_i$'s that destination two obtains during the first time slot.

Upon completion of the transmission strategy, destination one has access to 
\begin{align}
a_1, a_2, \ldots, a_{2m-2}, L_{2m+3}\left(\vec{a}\right), L_{2m+7}\left(\vec{a}\right).
\end{align}
Hence, receiver one has enough equations to recover its intended symbols. Similarly, destination two has access to 
\begin{align}
b_1, b_2, \ldots, a_{m-2}, L_{2m+4}(\vec{b}), L_{2m+6}(\vec{b}),
\end{align}
which allows destination two to recover its intended symbols.

\section{Concluding Remarks}
\label{Section:Conclusion}

We studied the DoF region of two-unicast layered wireless networks with delayed CSIT. We provided a set of new outer-bounds using the graph-theoretical notion of bottleneck nodes. We also provided networks in which these outer-bounds are tight and compared our results to prior work. We showed that unlike several recent DoF characterizations where the sum DoF only attain a small and finite set of values, the set of DoF values for two-unicast networks with delayed CSIT is in fact \emph{infinite}.

An interesting open problem is whether the set of values given in (\ref{Seq}), $\S$, includes all possible sum DoF for two-unicast layered networks with delayed CSIT. We already showed that our new outer-bounds do not suffice to characterize the DoF region. However, even for the example given in Section~\ref{Section:Discussion} the sum DoF is $3/2 \in \S$. Another interesting future direction is to study two-unicast networks with Shannon feedback rather than just channel state feedback.

\section*{Acknowledgment}

The author would like to thank Dr. Ilan Shomorony and Dr. Robert Calderbank for their insightful comments and hours of discussing this problem.

%
%
%


\bibliographystyle{ieeetr}
\bibliography{bib_TwoUnicastDelayed}

\begin{thebibliography}{10}

\bibitem{vahid2015informational}
A.~Vahid, I.~Shomorony, and R.~Calderbank, ``Informational bottlenecks in
  two-unicast wireless networks with delayed {CSIT},'' in {\em 53rd Annual
  Allerton Conference on Communication, Control, and Computing (Allerton)},
  pp.~1256--1263, IEEE, 2015.

\bibitem{FF56}
L.~R. Ford and D.~R. Fulkerson, ``Maximal flow through a network,'' {\em
  Canadian Journal of Mathematics}, vol.~8, pp.~399--404, 1956.

\bibitem{ACLY00}
R.~Ahlswede, N.~Cai, S.-Y.~R. Li, and R.~W. Yeung, ``Network information
  flow,'' {\em IEEE Transactions on Information Theory}, vol.~46,
  pp.~1204--1216, July 2000.

\bibitem{ADT11}
A.~S. Avestimehr, S.~Diggavi, and D.~Tse, ``Wireless network information flow:
  a deterministic approach,'' {\em IEEE Transactions on Information Theory},
  vol.~57, pp.~1872--1905, April 2011.

\bibitem{Ness2unicast}
C.~C. Wang and N.~B. Shroff, ``Beyond the butterfly: A graph-theoretic
  characterization of the feasibility of network coding with two simple unicast
  sessions.,'' {\em {In Proc. IEEE International Symposium on Information
  Theory}}, 2007.

\bibitem{ShenviDey}
S.~Shenvi and B.~K. Dey, ``A simple necessary and sufficient condition for the
  double unicast problem,'' {\em {in Proceedings of ICC}}, 2010.

\bibitem{xx}
T.~Gou, S.~Jafar, S.-W. Jeon, and S.-Y. Chung, ``Aligned interference
  neutralization and the degrees of freedom of the $2 \times 2 \times 2$
  interference channel,'' {\em {IEEE Trans. on Information Theory}}, vol.~58,
  pp.~4381--4395, July 2012.

\bibitem{WangTwoUnicast}
I.-H. Wang, S.~Kamath, and D.~N.~C. Tse, ``Two unicast information flows over
  linear deterministic networks,'' {\em Proc. of IEEE International Symposium
  on Information Theory}, 2011.

\bibitem{dof2unicastfull}
I.~Shomorony and A.~S. Avestimehr, ``Two-unicast wireless networks:
  Characterizing the degrees of freedom,'' {\em IEEE Transactions on
  Information Theory}, vol.~59, pp.~353--383, January 2013.

\bibitem{zeng2015alignment}
W.~Zeng, V.~R. Cadambe, and M.~M{\'e}dard, ``Alignment-based network coding for
  two-unicast-{Z} networks,'' {\em IEEE Transactions on Information Theory},
  vol.~62, no.~6, pp.~3183--3211, 2016.

\bibitem{TwoUnicastIsHard}
S.~Kamath, D.~N.~C. Tse, and C.-C. Wang, ``Two-unicast is hard,'' {\em Proc. of
  IEEE International Symposium on Information Theory}, July 2014.

\bibitem{CadambeJafar}
V.~R. Cadambe and S.~A. Jafar, ``Interference alignment and degrees of freedom
  for the {{K}}-user interference channel,'' {\em {IEEE Transactions on
  Information Theory}}, vol.~54, pp.~3425--3441, August 2008.

\bibitem{MotahariRealInterference}
A.~S. Motahari, S.~Oveis-Gharan, M.~A. Maddah-Ali, and A.~K. Khandani, ``Real
  interference alignment: Exploiting the potential of single antenna systems,''
  {\em IEEE Trans. on Information Theory}, vol.~60, pp.~4799--4810, August
  2014.

\bibitem{dofkkk}
I.~Shomorony and A.~S. Avestimehr, ``Degrees-of-freedom of two-hop wireless
  networks: ``everyone gets the entire cake'','' {\em IEEE Transactions on
  Information Theory}, vol.~60, pp.~2417--2431, May 2014.

\bibitem{WangTwoUnicastDCSIT}
I.-H. Wang and S.~Diggavi, ``On degrees of freedom of layered two unicast
  networks with delayed csit,'' {\em Proceedings of IEEE International
  Symposium on Information Theory}, July 2012.

\bibitem{XieTwoUnicastSecureDoF}
J.~Xie and S.~Ulukus, ``Sum secure degrees of freedom of two-unicast layered
  wireless networks,'' {\em IEEE Journal on Selected Areas in Communications},
  vol.~31, pp.~1931--1943, September 2013.

\bibitem{WangX2Unicast}
C.~Wang, T.~Gou, and S.~A. Jafar, ``Multiple unicast capacity of 2-source
  2-sink networks,'' {\em IEEE Global Telecommunications Conference}, 2011.

\bibitem{aggarwal2011achieving}
V.~Aggarwal, A.~S. Avestimehr, and A.~Sabharwal, ``On achieving local view
  capacity via maximal independent graph scheduling,'' {\em IEEE Transactions
  on Information Theory}, vol.~57, no.~5, pp.~2711--2729, 2011.

\bibitem{vahid2017interference}
A.~Vahid, V.~Aggarwal, A.~S. Avestimehr, and A.~Sabharwal, ``Interference
  management with mismatched partial channel state information,'' {\em EURASIP
  Journal on Wireless Communications and Networking}, vol.~2017, no.~1, p.~134,
  2017.

\bibitem{ilanthesis}
I.~Shomorony, {\em Fundamentals of multi-hop multi-flow wireless networks}.
\newblock PhD thesis, Cornell University, 2014.

\bibitem{ElGamal-Degraded}
A.~Gamal, ``The feedback capacity of degraded broadcast channels (corresp.),''
  {\em IEEE Transactions on Information Theory}, vol.~24, no.~3, pp.~379--381,
  1978.

\bibitem{el2011network}
A.~El~Gamal and Y.-H. Kim, {\em Network information theory}.
\newblock Cambridge university press, 2011.

\bibitem{davoodi2016aligned}
A.~G. Davoodi and S.~A. Jafar, ``Aligned image sets under channel uncertainty:
  {S}ettling conjectures on the collapse of degrees of freedom under finite
  precision {CSIT},'' {\em IEEE Transactions on Information Theory}, vol.~62,
  no.~10, pp.~5603--5618, 2016.

\end{thebibliography}

\end{document}